\documentclass[a4paper,fleqn]{cas-sc}

 \usepackage[numbers,sort&compress]{natbib}

\usepackage{amssymb}
\usepackage{amsmath,amsfonts}

\usepackage{algorithm}
\usepackage{algorithmic}

\usepackage{array}
\usepackage{multirow}
\usepackage{makecell}
\usepackage{adjustbox}
\usepackage{graphicx}
\usepackage{stfloats}

\usepackage{caption}
\usepackage{caption}
\usepackage{subcaption}
\usepackage{mwe}
\usepackage{comment}
\usepackage{url}
\usepackage{verbatim}
\usepackage{xcolor}
\usepackage{etoolbox}
\usepackage{xspace}
\newcommand{\vspacing}[2]{\rule[0pt]{0pt}{#1}\rule[-#2]{0pt}{0pt}}

\def\tsc#1{\csdef{#1}{\textsc{\lowercase{#1}}\xspace}}
\tsc{WGM}
\tsc{QE}

\begin{document}
\let\WriteBookmarks\relax
\def\floatpagepagefraction{1}
\def\textpagefraction{.001}

\shorttitle{Do Flow-Generation Tools Matter?}

\shortauthors{Rashed et al.}

\title[mode=title]{Do Flow-Generation Tools Matter? Unsupervised Characterization of MQTT Traffic Across Public Datasets}

\author[1]{Mohammed Rashed}[
orcid=0000-0002-0119-0796
]

\cormark[1]

\ead{mrashed@it.uc3m.es}

\credit{Conceptualization, Methodology, Software, Validation, Formal analysis, Investigation, Data curation, Writing -- original draft, Visualization}

\author[1]{Carlos García-Rubio}[
orcid=0000-0002-4635-722X
]

\ead{cgr@it.uc3m.es}


\credit{Conceptualization, Methodology, Supervision, Writing -- review and editing, Resources, Project administration, Funding acquisition}

\author[1]{Celeste Campo}[
orcid=0000-0003-1788-890X
]

\ead{celeste@it.uc3m.es}


\credit{Conceptualization, Methodology, Supervision, Project administration, Resources, Funding acquisition}

\affiliation[1]{
organization={Telematic Engineering Department, Universidad Carlos III de Madrid},
country={Spain}
}

\cortext[1]{Corresponding author}
\begin{abstract}
 Message Queue Telemetry Transport (MQTT) is widely used in IoT environments due to its lightweight publish-subscribe design and suitability for resource-constrained networks. While current MQTT security studies mainly focus on detecting attacks and optimizing classification performance, this work takes a different perspective by presenting a cross-dataset, tool-aware characterization of MQTT traffic at the flow level. This perspective is motivated by the fact that flow-generation is tool-dependent: flows are not directly observed in packet captures, but are constructed through tool-specific decisions about packet grouping, flow termination, and feature export.

Using packet captures from 4 public MQTT datasets, we generated over 380K flows with 3 open-access flow-generation tools: Zeek-flowmeter, Argus, and NTLFlowLyzer. We analyzed the tool-specific flows using K-means, DBSCAN, and HDBSCAN, with and without PCA-based dimensionality reduction. We measured clustering quality using 2 metrics: (1) Silhouette score to assess label-independent cluster compactness and separation, and (2) Adjusted Rand Index to measure agreement with available labels.

Our results show that the choice of flow-generation tool affects the resulting flow population, feature-space structure, and clustering behavior. Zeek-flowmeter achieves the best performance in several settings, particularly with K-means and PCA reduction to 10 components. Yet, no single tool consistently dominates across all datasets, algorithms, and evaluation metrics. These findings show that flow-level MQTT analysis should treat the dataset, flow-generation tool, preprocessing choices, clustering algorithm, and evaluation metric as interconnected experimental factors.
\end{abstract}

\begin{keywords}
MQTT \sep Network flows \sep Flow-generation tools \sep Unsupervised clustering \sep Intrusion detection \sep IoT security
\end{keywords}

\maketitle

\section{Introduction}
\label{sec:intro}
The Internet of Things (IoT) has become a major component of modern digital infrastructure, connecting heterogeneous devices across domestic, industrial, and critical-service environments. Recent estimates indicate that the number of connected IoT devices were to reach 21.1 billion by the end of 2025 and would continue growing toward 39 billion by 2030. This rapid expansion increases the importance of lightweight communication protocols that can operate efficiently under constrained bandwidth, processing, and energy conditions. Among these protocols, Message Queuing Telemetry Transport (MQTT) has become one of the most widely adopted messaging protocols for IoT communication because of its publish–subscribe architecture, low overhead, and suitability for unreliable or resource-constrained networks \cite{mqttver5}.

Its popularity stems from its ability to operate effectively in \textbf{\textit{constrained environments}} characterized by high latency, low bandwidth, and devices with limited processing power \cite{mqttHiveMQReport}. However, the same design principles that make MQTT efficient (minimalist headers and asynchronous nature \cite{mqttver5}), also make it a fertile ground for cyber-attacks. Traditional security perimeters often fail to account for the unique traffic patterns of MQTT, leaving IoT deployments vulnerable to sophisticated exploitation such as malformed packet injections, flood-based Denial of Service (DoS), and stealthy "slow-rate" DoS attacks that mimic legitimate telemetry \cite{husnain2022preventing,alzahrani2022artificial}. As MQTT continues to support increasingly diverse IoT deployments, understanding these behaviors at the network level becomes important for building effective monitoring and intrusion-detection mechanisms.

Existing MQTT security studies have mostly focused on detecting known attacks using supervised learning \cite{dikii2021attacks,hindy2020machine,khan2021deep}, deep learning \cite{syed2020denial,hindy2020machine}, and fuzzy logic \cite{ap2019secure}. While these approaches are useful for classification, they often rely on labeled datasets and are primarily designed to optimize detection performance rather than to characterize how benign and malicious MQTT traffic naturally organize under different traffic representations. Prior MQTT IDS studies already suggest that this representation is a consequential design choice. In particular, Hindy et al. \cite{hindy2020machine} reported that uniflow and biflow representations generally provide better discrimination for MQTT-related attacks than packet-level representations, and Khan et al. \cite{khan2021deep} similarly reported stronger and more stable results for flow-based features.

Flow-level analysis reformulates packet traces into connection-level records that capture temporal, directional, and volume-based properties of MQTT communication, rather than treating packets as isolated events. These aggregated properties are difficult to infer from individual packets alone and provide a broader view of communication behavior over time. The practical importance of flow-based analysis is evidenced by its common use in operational security monitoring and its adoption in commercial network security platforms that leverage flow telemetry to identify anomalous and malicious behavior \cite{ciscoStealth,arborNetScout}. However, MQTT traffic remains relatively underexplored in the literature from a flow-level characterization perspective compared with packet-level studies and supervised intrusion-detection approaches. This gap is central to our work: we seek not only to analyze MQTT traffic through clustering, but also to understand how different flow-generation tools influence the resulting representations and, consequently, the insights derived from them.
 
Public MQTT datasets differ substantially in scale, realism, and attack diversity, which already makes cross-dataset analysis challenging. In flow-level studies, however, there is an additional source of variability: the flow-generation process itself. A flow is not directly observed in a packet capture; it is constructed by a tool that decides how packets are grouped, when a flow starts and ends, which directionality model is used, and which features are exported. Consequently, the same PCAP can produce different flow populations depending on the extractor used, as we show in our study.

This tool dependence is important because changes in the generated flow population can alter feature distributions and, in turn, affect learning results. Evidence of this issue is provided by Haddadi and Zincir-Heywood \cite{haddadi2014benchmarking}, who benchmarked 5 open-source flow-generation tools using HTTP- and DNS-based botnet traffic and compared the resulting impact on three classification algorithms. Their results show that the choice of flow generator has significant influence on classifier performance. Therefore, flow-level MQTT analysis should not treat flow extraction as a neutral preprocessing step; it should consider the dataset, the learning algorithm, and the flow-generation tool as interconnected experimental factors.

To address this gap, this paper extends the tool-aware perspective of Haddadi and Zincir-Heywood to MQTT traffic through a cross-dataset, flow-level characterization. While Haddadi and Zincir-Heywood evaluated the effect of flow-generation tools through supervised classification, our work uses unsupervised clustering because the objective is characterization rather than detection. Instead of measuring how well a classifier predicts predefined labels, we examine how each tool shapes the MQTT feature space and whether the resulting flow representations form different grouping, separability, and label-agreement patterns. This perspective also motivates our use of both Silhouette score and ARI: Silhouette evaluates the compactness and separation of the clusters without relying on labels, whereas ARI measures the extent to which the resulting partitions agree with the original annotations.

We conduct this analysis using 3 open-access flow-generation tools: the Zeek-based \cite{zeekorg} Zeek-flowmeter \cite{zeekflowmeter}, Argus \cite{argusweb}, and NTLFlowLyzer \cite{shafi2025ntlflowlyzer}. We process packet captures from 4 publicly available MQTT datasets containing benign traffic and several MQTT-oriented attack scenarios \cite{ciklabakkal2019artemis,hindy2020machine,vaccari2020mqttset,poisson2024gothx}. From these captures, we generate more than 380K flow records across the 3 tools and analyze the resulting flow representations using K-means, DBSCAN, and HDBSCAN under different feature-space configurations, including experiments with and without PCA-based dimensionality reduction.

The main objective of this study is to examine how MQTT traffic is structured when represented as flows and how this structure changes across flow-generation tools. We evaluate the clustering outputs using two metrics: the Silhouette score, which measures cluster compactness and separation \cite{shahapure2020cluster}, and the Adjusted Rand Index (ARI), which measures the agreement between the generated clusters and the original dataset labels \cite{adjustedRand}. This allows us to compare both the internal quality of the clusters and their correspondence to known benign and attack categories.

Our results show that the choice of flow-generation tool has a substantial effect on the resulting flow population and clustering behavior. Zeek-flowmeter provides the strongest overall performance in several experimental settings, particularly when combined with K-means and PCA-based dimensionality reduction (10 dimensions). Nevertheless, the results also show that no single tool consistently dominates across all datasets, algorithms, and evaluation metrics. These findings suggest that MQTT flow analysis is not only dataset-dependent and algorithm-dependent, but also strongly tool-dependent.
 
The remainder of this paper is organized as follows. Section \ref{sec:bkgrd} provides background on MQTT communication and flow generation. Section \ref{sec:related} reviews related work on MQTT security, MQTT datasets, and flow-based analysis. Section \ref{sec:method} describes the dataset collection process, flow-generation tools, preprocessing steps, and clustering methodology. Section \ref{sec:res} presents the experimental results. Section \ref{sec:discussion} discusses implications of the results and the limitations of the study, and Section \ref{sec:conc} concludes the paper and outlines directions for future work.

\section{Background}
\label{sec:bkgrd}

\begin{figure}
    \centering
    \includegraphics[width=0.6\linewidth]{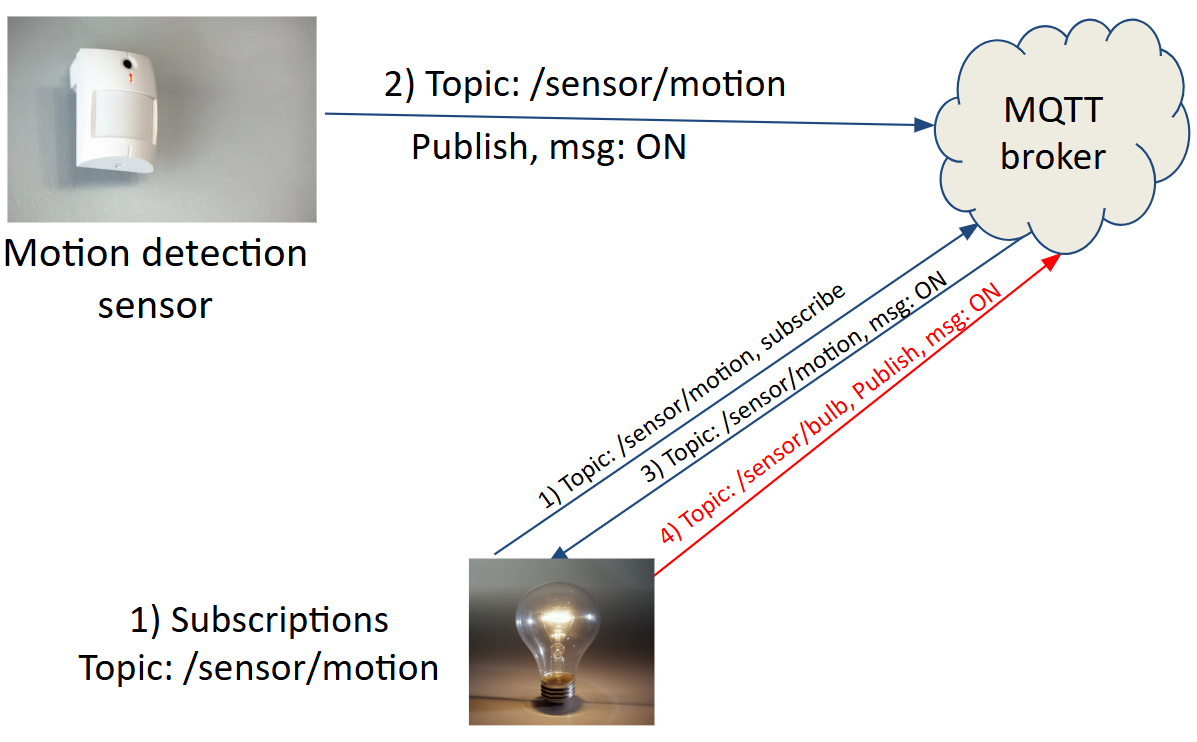}
    \caption{Example of a simple MQTT network with 2 MQTT-enabled devices, 2 topics and 1 MQTT broker}
    \label{fig:mqttNetEx}
\end{figure}

First introduced in 1999, MQTT has become one of the most widely used communication protocols in IoT environments. It follows a lightweight publish--subscribe model that minimizes network overhead and computational load, making it highly suitable for resource-constrained devices and unreliable or bandwidth-limited networks. MQTT operates on a client--broker architecture in which clients publish messages to specific topics, and the broker forwards these messages to all clients subscribed to those topics. This indirect communication model enables efficient message dissemination without requiring devices to maintain direct connections with each other \cite{amazQT}. Figure \ref{fig:mqttNetEx} provides an example  
where a motion sensor publishes messages to the topic \texttt{/sensor/motion} and a light bulb subscribes to that topic in order to react to motion events. The bulb would turn on when the motion sensor is activated (\texttt{ON}) and changes in the bulb status are then published 
on another topic: \texttt{/sensor/bulb}.

MQTT includes several features designed to improve reliability despite frequent disconnections and unstable connectivity, which are common in IoT deployments. One such feature is the \textit{Retain} flag, which instructs the broker to store the last published message on a topic and immediately deliver it to newly connecting subscribers. Another key feature is the \textit{Last Will and Testament (LWT)}, where a client specifies a message that the broker should publish automatically if the client disconnects unexpectedly. Together, these mechanisms reduce the perceived impact of intermittent connectivity and ensure that subscribers receive either the most recent state (through the retained message) or a failure notification (through LWT), thereby supporting more fault-tolerant communication \cite{eqmx}.

With regards to  Quality of Service (QoS), the MQTT protocol supports 3 levels: 0 (message sent at most once), 1 (message sent at least once), and 2 (message sent exactly once). These allow publishers to balance performance and reliability according to application needs. MQTT also maintains stateful client sessions, tracking subscriptions and storing undelivered messages, which is crucial for devices that frequently sleep or lose connectivity. For security, MQTT supports credential-based and certificate-based authentication, and encrypted communication through TLS/SSL. These capabilities collectively make MQTT a flexible and reliable protocol for IoT communication \cite{masteringMQTT,mqttessentials}.

While MQTT stands out for its simplicity, it is also vulnerable to several types of attacks. Some of these are \textit{protocol-specific}, exploiting MQTT message types or connection semantics. Examples include malformed packet attacks that target the broker's parser, SlowITe (a slow-rate Denial-of-Service -- \textit{DoS} -- attack that holds many connections open while sending packets slowly), and publish/subscribe abuse such as flooding the broker with excessive \texttt{CONNECT} or \texttt{PUBLISH} messages. Topic Hijacking, where unauthorized clients subscribe to or publish on restricted topics, is another MQTT-specific threat. Beyond MQTT-level threats, the protocol inherits all TCP-layer vulnerabilities since it operates over TCP. These include SYN flooding, TCP reset attacks, and other forms of connection exploitation \cite{mqttattacktypes,dadkhah2024ciciomt2024,hindy2020machine,vaccari2020slowite,dachmanTCP2019}. This combination of application-layer and transport-layer behavior motivates the use of flow-level analysis, since attacks may be reflected not only in individual packet contents but also in connection duration, packet timing, directionality, and traffic volume.

\subsection{Network Flow}
\label{sec:bkgrd:flow}
Network flows provide a reliable mechanism in the identification and categorization of traffic, and enable administrators in pinpointing unauthorized activities and threats over the network \cite{rashmi1}. A network flow is essentially a set of packets that share common attributes/properties. Any given set is typically identified using a 5-tuple key, in which the packets of the set share \texttt{source IP, source port, destination IP, destination port, protocol}
. In the real-world, the flow key could also include the VLAN ID among other attributes. A flow groups packets using a common key, which is consequently used to classify traffic\footnote{similar to running the \texttt{netstat -na} command}. Flows use counters in order to keep track of packets within the flow as well as additional attributes e.g. statistics, flow timers, etc. \cite{ntop1}.

A flow cache, typically built using a hash table, is the data structure used to keep network flows. Flow caches are continuously fed with arriving packets, and they store the active flows in memory. The first arriving packet is considered the start of a flow. Typically, a flow that does not have a normal TCP connection teardown, ends once any of the two aging timers expires; this happens when a flow is idle for too long or when a flow is stored in the cache for too long \cite{ntop1}. This means that the same packet capture can produce different flow records depending on the tool used and the timeout or export rules applied by that tool. Flow-generation tools may differ in whether they preserve one long-lived flow as such, split it into multiple records, or terminate it after an inactivity threshold. 
In the context of our study, it is also important to differentiate between packets, MQTT sessions, and flow records. While MQTT defines application-level interactions between clients and brokers, flow-generation tools operate mainly at the network and transport layers. 
Thus, the three tools used in this study do not reconstruct MQTT application sessions as such. Instead, they generate flow records primarily from network- and transport-layer observations, such as addresses, ports, protocol state, packet counts, byte counts, and timing behavior. Therefore, the exported records should be interpreted as connection-level summaries of TCP-based MQTT traffic rather than as direct representations of MQTT publish/subscribe interactions. 

\subsection{Flow-Generation Tools}
\label{sec:bkgrd:tools}
Our work involves the use of 3 different flow-generation tools, namely: Argus, NTLFlowLyzer and Zeek-based zeek-flowmeter. Argus \cite{argusweb} is a project started at the Software Engineering Institute of Carnegie Mellon University, USA; it was first released to the public in the beginning of 1996. The current version, which we used in our analysis is Argus 3.0. With regards to NTLFlowLyzer, it is a flow-generation tool that was developed by researchers from the Canadian Institute for Cybersecurity (CIC). This python-based tool seeks to resolve the issues that were found in the CICFlowMeter \cite{gil2016characterization} developed earlier by the same research group as well as add more descriptive fields of the flows. With regards to Zeek-flowmeter \cite{zeekflowmeter}, which relies fully on Zeek's engine \cite{zeekorg} for flow generation, it partially imitates the fields of CICFlowMeter (hence the name similarity) using a simple add-on script. We selected these 3 tools because they are open-source, sufficiently documented for reproducible PCAP-to-flow conversion, and \textbf{most importantly due to their representation of different flow-construction approaches}. Argus provides transaction-oriented flow reporting with periodic status exports, NTLFlowLyzer produces a large bidirectional statistical feature set, and Zeek-flowmeter inherits Zeek’s connection-state logic. This combination makes them suitable for a tool-aware comparison of how flow-generation semantics affect downstream MQTT clustering. In this study, the tool-specific flow-generation semantics are central to interpreting both the number of generated flows and the behavior of the clustering algorithms applied later in the analysis pipeline. 

\subsubsection{Argus}
\label{sec:tools:argus}
Argus reports discovered network transactions as flow activity status records. By default, each Argus record contains a single pair of timestamps corresponding to the first and last packet observed for a given flow during the observation period. For long-lived flows, Argus periodically reports flow activity every \texttt{ARGUS\_FLOW\_STATUS\_INTERVAL} seconds, provided that new activity has been observed on the flow; the default interval is 60 seconds \cite{argusServerDebian,argusRscFile}.

\subsubsection{NTLFlowLyzer}
\label{sec:tools:lyzer}

NTLFlowLyzer generates bidirectional flows from network- and transport-layer traffic. The first packet of a flow determines the forward direction, defined as source to destination, while the reverse direction is defined as destination to source. For TCP traffic, flows are terminated when the connection is torn down by a FIN or RST packet, when the flow exceeds the configured maximum duration, or when the flow remains inactive beyond the configured timeout. The default value of \texttt{max\_flow\_duration} is 120,000 seconds; when this threshold is exceeded, the current flow is terminated and a new flow is initiated. The default value of \texttt{activity\_timeout} is 5,000 seconds; if no packet is observed for a flow during this interval, the flow is terminated due to inactivity \cite{lyzerGithub}.

\subsubsection{Zeek}
\label{sec:tools:zeek}
 Zeek represents TCP traffic as connection records. Each connection is identified by the communicating endpoints and transport protocol, i.e., the source and destination IP addresses, source and destination ports, and protocol. The connection start time corresponds to the timestamp of the first observed packet. The connection duration represents the time for which the connection lasted; however, for TCP, Zeek excludes trailing non-productive packets after one direction has closed, such as the final ACK in a regular 3-way/4-way TCP teardown. Zeek further records the observed TCP state, including connection attempts, established but unterminated sessions, normal terminations, resets, half-open connections, and midstream traffic for which no SYN was observed. In the absence of further activity, Zeek applies a TCP inactivity timeout, which is set to 5 minutes by default \cite{zkDocProtocols,zkDocBare}. Zeek-flowmeter does not define an independent TCP flow identity; instead, flow identification is delegated to the underlying Zeek engine, while Zeek-flowmeter performs layer 3/4 feature extraction on the resulting Zeek flows \cite{zeekflowmeter}.

\section{Related Work}
\label{sec:related}

\subsection{MQTT Security and Attacks} 
As an application layer protocol, devices that run MQTT may suffer vulnerabilities that require certain countermeasures.  Husnain et al. \cite{husnain2022preventing} 
found 81 publicly disclosed vulnerabilities\footnote{Vulnerabilities were reported in the National Vulnerability Database (NVD) and the Common Vulnerabilities and Exposures (CVE) program; see \url{https://nvd.nist.gov} and \url{https://www.cve.org/}.} between April 2014 and December 2021. They classified them into 3 main categories:
\begin{itemize}
\item \textit{Required Fields Checks} (37 vulnerabilites): Because the required fields in MQTT packets depend on the packet type, improper validation (e.g. omitting password check when a username is supplied) was shown to cause server crashes, buffer overflows, and remote code execution.
\item \textit{Lack of Logical Errors Checks} (25): Missing logical error checks in MQTT networks can lead to crashes, buffer overflows, and DoS attacks when processing crafted packets. The reported vulnerabilities show exploits of fields like \textit{Keep-Alive} and \textit{Will Delay Interval}.
\item \textit{Improper Packet Length Checks} (12): Packet parsing helps identify packet fields. Yet, without proper length checks, attackers can exploit sequential parsing using crafted packets. This can cause remote code executions, buffer overflows, DoS, and unauthorized memory access.
\end{itemize} 
Based on this analysis, Husnain et al. propose an MQTT packet-parsing engine that integrates with MQTT network-based IDS and runs extensive validation checks on 
packet fields. The engine would implement these 5 steps: \textit{1) protocol intuition, 2) vulnerability assessment, 3) protocol identification, 4) parsing}, and \textit{5) strict validation}.

Hintaw et al. \cite{hintaw2019brief} studied  MQTT's security challenges and solutions by analyzing security considerations at different layers of IoT systems that use MQTT 
mechanisms such as IPSec or Host Identity Protocol (HIP) can provide confidentiality and authentication, but they introduce additional overhead and deployment complexity.
For the Network layer, they proposed using IPSec for confidentiality and authentication but noted the limitation of extra headers, with 6LowPAN compression being a possible solution to this limitation. An alternative approach would be the Host Identity Protocol (HIP) \cite{rfc7401} as it would separate node identifiers and brokers from their locators
. At the Transport layer, using TLS Session Resumption can notably reduce repeated TLS Handshakes although implementing TLS is not practical in this resource-constrained context. An alternative involves using hardware-based TLS implementations despite being costly in terms of hardware and power consumption. Generally, the use of certificates and TLS is recommended in MQTT's specification whenever possible \cite{mqttver5}. The use of TLS helps protect traffic, but incorrect TLS configuration (e.g. vulnerable TLS versions used for compatibility reasons with legacy devices) would make the network exposed.
At the application layer, \textit{OAuth} or \textit{LDAP} (both external authentication systems) can be used for brokers. This external system would be responsible for enabling access to tokens and for credential-based authentication, but would involve complexities such as latency. Hintaw et al. consider several drawbacks to current MQTT security mechanisms. These include high complexity emerging from the suggested approaches, which requires using more resources. Also, while these proposed measures may solve challenges, they also may introduce new vulnerabilities in the process. Finally, the need for additional services is another drawback. When devices have sufficient resources, traditional encryption methods for MQTT traffic at the application layer are viable. This is not the case of these resource-limited devices.

\subsection{Supervised Intrusion Detection for MQTT}

Most MQTT intrusion detection research follows a supervised learning paradigm. A foundational contribution in this area is the work of Hindy et al.~\cite{hindy2020machine}, who introduced the MQTT-IoT-IDS2020 dataset and evaluated several ML classifiers across packet-level, uniflow, and biflow traffic representations. Their results showed that flow-based representations generally provide better discrimination for MQTT-related attacks, whereas packet-level representations can be sufficient for more traditional networking attacks. This finding highlights that traffic representation is not merely a preprocessing choice, but a fundamental methodological decision in MQTT intrusion detection.

Building on this observation, Khan et al.~\cite{khan2021deep} proposed a Deep-Learning (DL)-based IDS and compared its performance against Naive Bayes (NB), Random Forest (RF), K-Nearest Neighbors (KNN), Decision Tree (DT) using MQTT-IoT-IDS 2020's dataset for the experiments. While the study's objective is to demonstrate the improved performance of their proposed DL-based solution vs other ML-based ones, the results also reflect a much better overall performance for the Uni-Flow and Bi-Flow over the packet-based features. With a few exceptions, the packet-based analysis is generally outperformed by both Uni-Flow and Bi-Flow-based analyses.
Also, the range of performance results is wider for packet-based experiments. For example, packet-based accuracy values' range is (49.8, 94.9), while it is (94.2, 99.8) for Bi-Flow and (90.1, 99.1) for Uni-Flow consecutively. 
These results show the clear advantages of flow-based representations over packet-level features in intrusion detection.

Another study that focused on improving classification accuracy is the work of Alzahrani and Aldhyani \cite{alzahrani2022artificial}. The authors tested several traditional ML algorithms: \textit{KNN}
and \textit{Linear Discriminant Analysis (LDA)} as well as DL models: \textit{Convolutional Neural Network (CNN)} and \textit{CNN Long Short-Term Memory Networks (CNN-LSTM)} on the MQTTset dataset and achieved high classification performance; accuracy reached as high as 98.9\% for CNN-LSTM. While using a multi-class dataset, the authors applied binary classification, where they gathered all attack traffic together vs. normal traffic. 
While useful as targeted evaluations, this class of studies contribute relatively little to broader questions regarding traffic abstraction, feature generation, or generalization across datasets.

\subsection{Unsupervised and Anomaly-Based MQTT IDS}
 
Unsupervised and anomaly-based MQTT IDSs are particularly relevant because they reduce dependence on attack labels and offer the potential to identify previously unseen threats, reflecting a real-world perspective. ARTEMIS, whose dataset is used in our experiments, was among the earlier efforts in this direction, proposing a lightweight anomaly-based IDS for MQTT environments using packet-level features and several anomaly detection algorithms~\cite{ciklabakkal2019artemis}. In the experiments, the authors tested a mix of supervised (e.g. RF) and unsupervised algorithms (e.g. K-means) on the 2-class dataset. Although promising, ARTEMIS remained focused on packet-level analysis and did not investigate flow-based representations.

Ranjalahy et al.~\cite{ranjalahy2025hybrid} introduced a more recent work that explored representation learning approaches. The VAE+KNN framework learns a latent representation of benign MQTT behavior using a Variational Auto-Encoder (VAE) and subsequently measures anomaly scores through KNN distances in the latent space. Evaluated on MQTTset-derived benign and attack traffic data, and benign traffic data from MQTT-IoT-IDS2020, the authors reported improvements over One-Class Support Vector Machine (OCSVM) and Isolation Forest baselines. However, their approach is also more costly and less suitable than the baseline algorithms.

GAN-AE extends this idea through adversarial representation learning. This work, by  Boppana and Bagade \cite{boppana2023gan}, relies on Generative Adversarial Networks' ability in modeling  complex and high-dimensional patterns in provided data. GAN-based training on an
autoencoder (AE) model can help boost the model's ability to identify and detect anomalies. This work's anomaly detection model is trained exclusively on normal MQTT traffic. It then uses this knowledge to detect anomalies through reconstruction error~\cite{boppana2023gan}. Importantly, besides using Uni-Flow features from MQTT-IoT-IDS2020, the authors employed Tranalyzer to generate flows from their own lab-setup, making it one of the few MQTT IDS studies that documents a flow-generation process. Their results show a clear improvement of their proposed GAN-AE's performance compared to just using an Enocder. However, 
the work remains primarily concerned with its detector performance rather than understanding how alternative flow-generation tools might alter the structure of the underlying dataset. 

\subsection{Flow Generation and Tool Dependence}

As noted from the earlier mentioned works, network-based intrusion and anomaly detection commonly relies on either packet-based or flow-based traffic representations. Compared with packet captures, flow records provide a more aggregated view of network connections, typically omitting payloads and retaining metadata such as endpoints, ports, protocol, duration, packet counts, byte counts, and TCP flags~\cite{ring2019survey}. In the context of IDS and anomaly detection research, this compact representation makes flow data attractive for scalable traffic analysis compared with packet-level inspection. Yet, it is worth noting that flow extraction is not a neutral preprocessing step. 
In practice, a network flow is not a naturally occurring object, but a tool-dependent abstraction whose properties depend on the inner design and mechanisms that the tool itself has. In the context of HTTP and DNS-based botnets, Haddadi and Zincir-Heywood \cite{haddadi2014benchmarking} studied the differences between 5 flow generation tools showing that not all of them generated all the feature categories. Particularly, 2 of their 5 tools did not generate any Inter-Arrival features. 
The authors highlighted that, in some cases, a tool might not generate any flows due to its predefined rules~\cite{haddadi2014benchmarking}.
Within the MQTT security literature, flow generation received little attention. Hindy et al.~\cite{hindy2020machine} compared packet, uniflow, and biflow representations, but did not treat the flow-generation tool itself as an experimental variable. Similarly, GAN-AE explicitly specified Tranalyzer as its flow extractor, but did not investigate how alternative tools might produce different flow populations from the same traffic traces~\cite{boppana2023gan}. 

\subsection{Unsupervised Clustering for Traffic Characterization}

Unsupervised clustering has been used in network traffic analysis for tasks such as traffic classification and anomaly detection. For example, Erman et al. \cite{erman2006traffic} used K-means and DBSCAN to classify network traffic protocols. Similarly, 
M\"{u}nz et al. \cite{munz2007traffic} proposed an anomaly detection approach using K-means clustering over flow records. Their method aggregates flows into service-specific time intervals and clusters traffic using features such as packet counts, byte counts, and IP/port source-destination pairs. Their time-based experiment on lab-generated traffic showed that their clustering approach was successful. Their results show high similarity between the clusters formed for both normal and anomalous attacks and their generated traffic. 

These works show that clustering can be useful when labels are incomplete or when the goal is to expose behavioral structure in traffic. However, they generally treat the traffic representation as fixed. In contrast, our work uses clustering not as a detector or label generator, but as an analytical lens to compare how different flow-generation tools shape the structure of MQTT traffic.

Within MQTT research, clustering has generally been used as an auxiliary component within IDS pipelines rather than as a primary mechanism for understanding the structure of traffic itself. Consequently, little attention has been paid to how clustering behavior changes across datasets or across different flow-generation methodologies. This is particularly important in flow-based analysis, where the observed feature space is shaped not only by the underlying traffic but also by the extraction tool used to transform packets into flow records.

\subsection{Research Gap}

The existing literature establishes three important observations. First, MQTT requires protocol-aware datasets and security mechanisms~\cite{vaccari2020mqttset,hindy2020machine,ciklabakkal2019artemis}. Second, traffic abstraction matters, and flow-based representations frequently outperform packet-level approaches in MQTT intrusion detection~\cite{hindy2020machine,khan2021deep}. Third, unsupervised and anomaly-based methods offer promising capabilities for identifying previously unseen attacks~\cite{boppana2023gan,ranjalahy2025hybrid}.

Despite these advances, an important gap remains. Existing studies typically treat flow extraction as a fixed preprocessing stage and evaluate IDS performance using a single dataset and a single flow representation. As a result, little is known about how different flow-generation tools influence the resulting flow population and, consequently, the behavior of downstream learning algorithms. This work addresses that gap. Rather than proposing yet another MQTT IDS, we perform a cross-dataset, tool-aware, unsupervised characterization of MQTT traffic. By comparing Zeek-flowmeter, Argus, and NTLFlowLyzer across multiple MQTT datasets and evaluating their impact on clustering outcomes, we investigate how the process of transforming packets into flows fundamentally shapes the observable structure of MQTT network traffic.

\section{Methodology}
\label{sec:method}

\subsection{Scope of the Study}
\label{sec:method:scope}
This study evaluates how different flow-generation tools affect the unsupervised characterization of MQTT traffic. The experimental pipeline consists of 3 main blocks: \textit{1) dataset selection and preparation}, \textit{2) flow generation and cleansing} and \textit{3) clustering and evaluation}.  
In order to provide a real-world approach, we avoid changing the original class distributions whenever possible. In all cases, we maintain the original percentages of the different classes within our datasets; we avoid any balancing techniques such as oversampling, undersampling, SMOTE, etc. With regards to the flow generation tools, we use them in the very basic settings and only introduce necessary additions/modifications to the default configuration. Any balancing to the datasets and configurations beyond default settings of flow generation tools are beyond the scope of the study and left for future work. 

\begin{figure}
    \centering
    \includegraphics[width=\linewidth]{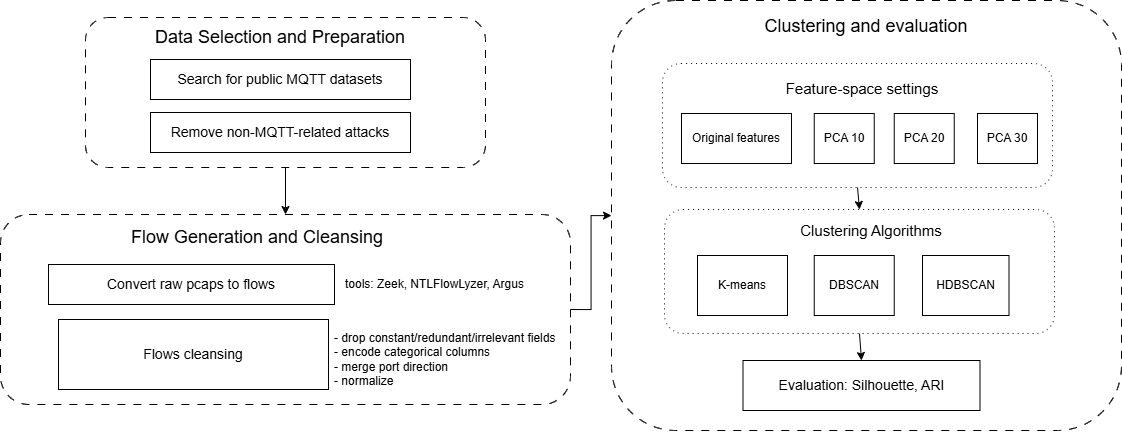}
    \caption{Pipeline for Experiments} 
    \label{fig:pipeline}
\end{figure}

Also, encrypted MQTT traffic (MQTTS) is outside the scope of this study. We intentionally focus on unencrypted MQTT first in order to isolate the effect of flow-generation tools without introducing encryption as an additional source of variability. Because the selected tools operate primarily on network-layer and transport-layer behaviors, some tool-dependent effects may also persist under encryption; however, evaluating that setting requires a separate analysis of TLS session behavior and encrypted traffic visibility. We therefore treat unencrypted MQTT as a controlled first step and leave MQTTS for future work. 
This section details the dataset collection and flow generation and the workflow that we designed to analyze our MQTT network traffic dataset (benign and malicious). An overview of our pipeline is provided in Figure \ref{fig:pipeline}. 

\subsection{Dataset Collection}
\label{sec:datasets}
In this work, we sought an approach that would help us build a bird-eye's view of existing MQTT attacks based on publicly available datasets. To this end, we carried out an exhaustive search of datasets that were addressed in the literature and that focused on MQTT attacks. While several datasets claim that they involve MQTT attacks, this does not necessarily fit our criteria. That is because some datasets are generated in the context of an MQTT network, but the attacks are not focused on the application level -- where the MQTT protocol actually sits. Thus, our dataset collection process involved the following phases:
\begin{enumerate}
    \item \textit{Literature Search}: We searched the literature for mentions of MQTT attack-based datasets.
    \item \textit{Analysis of the descriptions of the datasets}: we discarded any dataset that did not involve attacks on the MQTT protocol itself. If the dataset included mixed attack types, we discarded the data points that represent attacks that were not on MQTT.
\end{enumerate}

Using this approach, we ended up selecting a collection of 4 datasets that includes benign traffic as well as multiple MQTT-oriented attack scenarios, including flooding denial-of-service, brute-force authentication, malformed data, publish flooding, and SlowITe-style slow-rate attacks. Our final collection contains 1.74M benign raw packets and 10.74M malicious packets. The selected datasets differ in scale, testbed design, traffic-generation tools, and attack diversity. Table \ref{tab:mqtt-datasets} summarizes these datasets, including the year of publication, whether the traffic was generated through simulation or real testbeds, the application domain, the attack types, the number of benign and malicious packets, and the tools used to generate malicious traffic.

\renewcommand{\arraystretch}{1.4} 
\begin{table*}[ht]
\centering
\resizebox{\textwidth}{!}{%
\begin{tabular}{cccccccccc}
\hline 
\multirow{2}{*}{Dataset} & \multirow{2}{*}{Year} & \multicolumn{2}{c}{Traffic} & \multirow{2}{*}{Testbed} & \multirow{2}{*}{Domain} & \multirow{2}{*}{Attack} & \multirow{2}{*}{Packets} & \multirow{2}{*}{Tool}& \multirow{2}{*}{\begin{makecell}[t]{
Benign\\Pkts.}
\end{makecell}}\tabularnewline
\cline{3-4}
 &  & Sim. & Real &  &  &  &  & \tabularnewline
\hline 
Artemis & 2019 & Y & Y & N/A & - & Flooding: DoS  & 422K & MQTT-malaria & 179K\\
\cite{ciklabakkal2019artemis}& & & &&&&&&\tabularnewline
\hline 
\multirow{5}{*}{MQTTset} & \multirow{5}{*}{2020} & \multirow{5}{*}{Y} & \multirow{5}{*}{} & \multirow{5}{*}{10 sensors} & \multirow{5}{*}{\begin{makecell}[t]
\centering
Smart Home
/Office
\end{makecell}} & Flooding: DoS  & 130K & MQTT-malaria & \multirow{3}{*}{}\tabularnewline
\cline{7-9}
 &  &  &  &  &  & MQTT Publish Flood & 613 & IoT-Flock \tabularnewline
\cline{7-9}
 &  &  &  &  &  & Brute-Force Authentication & 14K & MQTTSA & 11M \tabularnewline
\cline{7-9}
\cite{vaccari2020mqttset} &  &  &  &  &  & malformed data & 10K & MQTTSA & (discarded)\tabularnewline
\cline{7-9}
 &  &  &  &  &  & SlowITe & 9.2K & SlowITe\tabularnewline
\hline 
\multirow{2}{*}{\begin{makecell}[t]{
MQTT-IoT-IDS2020}
\end{makecell}} 
& \multirow{2}{*}{2020} & \multirow{2}{*}{Y} & \multirow{2}{*}{} & 12 sensors/ & \multirow{2}{*}{\begin{makecell}[t]
\centering
Smart Home 
\end{makecell}} & \multirow{2}{*}{MQTT brute-force} & \multirow{2}{*}{10M} & \multirow{2}{*}{MQTT-PWN} & \multirow{2}{*}{1.07M}\tabularnewline
\cite{hindy2020machine} &  &  &  & sim. cam. feed &  &  &  & \tabularnewline
\hline 
\multirow{5}{*}{\begin{makecell}[t]{
GothX -MQTTset\\
Reproduction}
\end{makecell}} & \multirow{5}{*}{2024} & \multirow{5}{*}{Y} & \multirow{5}{*}{} & \multirow{5}{*}{10 sensors} & \multirow{5}{*}{-} & Malformed data & 1.09K & MQTTSA & \multirow{5}{*}{494K} \tabularnewline
\cline{7-9}
 &  &  &  &  &  & SlowITe & 31K & SlowITe\tabularnewline
\cline{7-9}
 &  &  &  &  &  & brute force authentication & 21K & MQTTSA\tabularnewline
\cline{7-9}
\cite{poisson2024gothx} &  &  &  &  &  & MQTT Publish Flood & 7.29K & IoT-Flock \tabularnewline
\cline{7-9}
 &  &  &  &  &  & Flooding Denial of Service  & 90K & MQTT-malaria\tabularnewline
\hline 
\end{tabular}
}
\smallskip
\caption{Overview of MQTT-based IoT datasets, traffic generation setup, attack scenarios, num. packets and tools for generating malicious traffic.}
\label{tab:mqtt-datasets}
\end{table*}

\subsection{Flow Generation and MQTT Filtering}
\label{sec:methods:flow}

After selecting the datasets, we converted the packet captures into flow records using the 3 flow-generation tools: 1) Argus \cite{argusweb}, 2) NTLFlowLyzer \cite{shafi2025ntlflowlyzer} and 3) The Zeek-based \cite{zeekorg} flow extractor known as Zeek-flowmeter (for simplicity, we will call it Zeek from now onwards) \cite{zeekflowmeter}. Since the objective of the study is to compare the outputs of these tools under the same input conditions, we retained only packet captures that could be successfully processed by all three tools. If any tool failed to generate flows for a given packet capture, that capture was excluded from the final analysis.

\begin{figure}
\centering
\includegraphics[width= \textwidth]{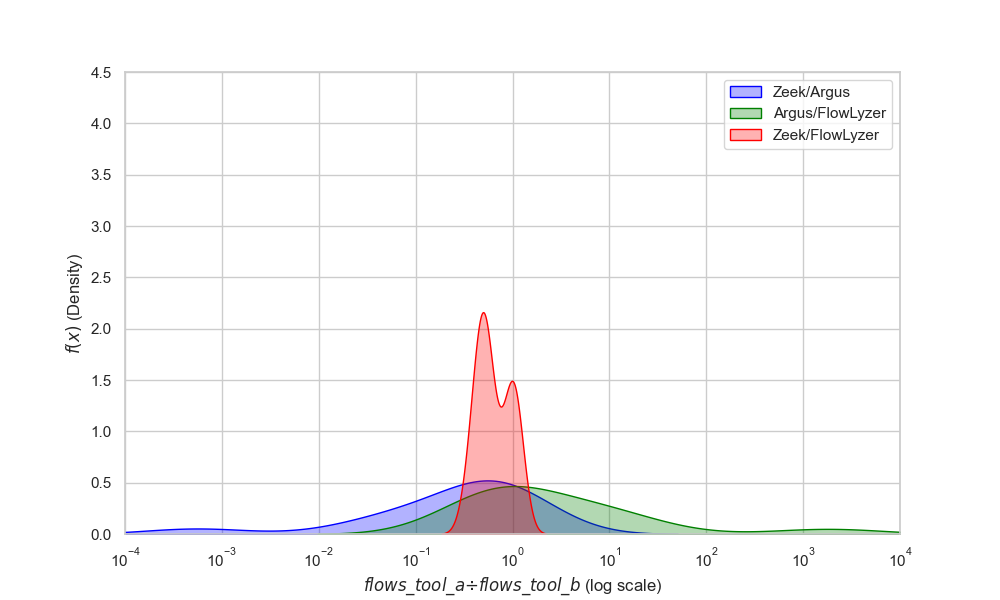}
\caption{Probability Density Function of ratios between the number of flows for each pair of tools}
\label{fig:pdfsFlows}
 \end{figure}

This exclusion step resulted in 14 packet-capture files that were successfully processed by all the 3 tools. Table \ref{tab:flows} reports the number of generated flows for each dataset, capture file, and flow-generation tool. Across the retained captures, the 3 tools produced a total of 380,575 flow records. The number of generated flows varied substantially across tools, indicating that flow generation is clearly not a neutral preprocessing step. Instead, tool-specific flow identification, timeout, and export mechanisms directly influence the size and structure of the resulting flow datasets as we show later. Figure \ref{fig:pdfsFlows} shows the Probability Density Function (PDF) for the pairwise ratio between the flows generated by each tool:
\(\frac{Zeek}{Argus}\),\(\frac{Argus}{NTLFlowLyzer}\) and \(\frac{Zeek}{NTLFlowLyzer}\). 
The figure shows that the Zeek/NTLFlowLyzer ratio is mostly concentrated around approximately $10^{-0.3}$ with less spread, suggesting that Zeek tends to generate 0.5 of the flow counts that NTLFlowLyzer generates for any given PCAP file. 
By contrast, the broader Argus-based ratios indicate greater variability in the number of records exported from the same captures.

To ensure that the analysis focused on MQTT traffic, we filtered the generated flows by MQTT unencrypted traffic port 1883. A flow was retained if either the source or destination port corresponded to port 1883. This filtering step removed non-MQTT flows and ensured that the clustering experiments were applied only to traffic associated with MQTT communication. Since the tools differ in the number and type of fields they generate, all available fields were initially exported from each tool before preprocessing. The number of fields for each tool vary. For instance, while Zeek and Argus can generate a maximum of 79 fields and 84 fields consecutively, NTLFlowLyzer can generate up to 348 fields. We modified Zeek to add the source and destination ports to the fields making the total number of fields 81. For MQTT-IoT-IDS2020, we generated the complete set of flows for all tools, but used a proportional subset in the clustering pipeline due to the large size of the resulting flow datasets. The subset was selected separately for each tool and class. Specifically, for each CSV file representing a labeled class, we retained the first (n\_c) flow records in their generated order, where (n\_c) was chosen to preserve the original class ratio observed in the corresponding tool-generated flow CSV file.

\subsection{Feature Preprocessing}
\label{sec:methods:features}
The generated flow datasets were then preprocessed to improve consistency and reduce noise. Columns with a single unique value were removed because they do not contribute to clustering. Fields that were irrelevant to behavioral characterization or could create artificial separation, such as explicit start and end timestamps, were also excluded. Where applicable, categorical fields were encoded numerically, and numeric values were downcast from larger to smaller data types to reduce memory usage. In addition, source and destination port information was transformed into a direction-related feature indicating whether the flow was directed to or from port 1883.

Additional preprocessing was applied to NTLFlowLyzer because it generated a substantially larger number of features than the other tools as explained in Section \ref{sec:methods:flow}. 
To reduce redundancy, statistical features derived from already retained base fields were removed when the original fields were available. When the underlying base feature was not available, summary statistics such as the mean and standard deviation were retained. This step was intended to reduce feature duplication while preserving relevant behavioral information.

\begin{table}[!hbtp]
    \centering
    \small
    \begin{tabular}{llrrrr}
        \hline
        \textbf{Dataset} & \textbf{File} & \textbf{Zeek} & \textbf{Argus} & \textbf{Flowlyzer} & \textbf{TotSetFlws} \\ 
        \hline
        & slowite & 1,025 & 6,026 & 1,025 & \\
        & normal & 30,012 & 30,012 & 59,586 & \\
      GothX\_MQTTset  & bruteforce & 2,053 & 2,063 & 4,078 & \\
       \_Repro & malformed & 24 & 36 & 65 & \\
        & flood & 2 & 4 & 3 & \\
        & malaria & 16 & 311 & 31 & \\
        \cline{2-6}
        & \textbf{TOTAL} & \textbf{33,132} & \textbf{38,452} & \textbf{64,788} & \textbf{136,372} \\ 
        \hline
        \multirow{4}{*}{MQTTset\_Original} 
        & malariaDoS & 16 & 499 & 31 & \\
        & malformed & 903 & 1,849 & 2,250 & \\
        & bruteforce & 1,431 & 2,875 & 2,861 & \\
        & slowite & 1,025 & 2,049 & 1,025 & \\
        \cline{2-6}
        & \textbf{TOTAL} & \textbf{3,375} & \textbf{7,272} & \textbf{6,167} & \textbf{16,814} \\ 
        \hline
        \multirow{2}{*}{MQTT-IoT-IDS2020} 
        & bruteforce & 70,001 & 54,106 & 63,351 & \\
        & normal & 5,585 & 5,585 & 11,169 & \\
        \cline{2-6}
        & \textbf{TOTAL} & \textbf{75,586} & \textbf{59,691} & \textbf{74,520} & \textbf{209,797} \\ 
        \hline
        \multirow{2}{*}{artemis} 
        & attack & 1,373 & 11,211 & 1,373 & \\
        & benign & 2 & 3,631 & 2 & \\
        \cline{2-6}
        & \textbf{TOTAL} & \textbf{1,375} & \textbf{14,842} & \textbf{1,375} & \textbf{17,592} \\ 
        \hline
        \multicolumn{2}{l}{\textbf{TOT.ToolFlows}} & \textbf{113,468} & \textbf{120,257} & \textbf{146,850} & \textbf{380,575} \\ 
        \hline
    \end{tabular}
    \caption{No. of flows generated per dataset and per tool. The total no. of generated flows by the 3 tools is over 380K.}
    \label{tab:flows}
\end{table}

\subsection{Clustering Configuration}
\label{sec:clustering}
To ensure a robust and comprehensive evaluation, this study adopts a multi-algorithm comparative framework that aligns with recent advances in network anomaly detection. Contemporary research highlights the importance of evaluating both statistical and density-based clustering approaches to effectively model the heterogeneous nature of network traffic. For example, Baklashov and Kulyabov \cite{baklashov2025statistical} demonstrated the value of comparing statistical baseline methods with density-based clustering techniques in modern network environments. Similarly, Diallo and Patras \cite{diallo2021adaptive} showed that the severe class imbalance commonly observed in network traffic can significantly reduce the effectiveness of traditional supervised classification methods. In contrast, clustering approaches can adapt more naturally to such distributions by identifying minority attack patterns without requiring balanced class labels. Given that real-world network traffic is characterized by extreme class imbalance, complex spatial structures, and high-dimensional feature spaces, our methodology employs a three-stage clustering pipeline consisting of K-Means, DBSCAN, and HDBSCAN. This progression enables the evaluation of centroid-based, density-based, and hierarchical density-based clustering paradigms under consistent experimental conditions. The hyperparameter ranges were selected to explore a diverse set of clustering configurations without incurring excessive computational cost, while remaining suitable for the characteristics of the evaluated datasets.

The clustering experiments were conducted separately for each dataset and each flow-generation tool. This design avoids merging datasets with different traffic-generation conditions and allows the effect of each tool to be evaluated within the same dataset context. For each tool–dataset pair, we tested our 3 clustering algorithms.

\subsubsection{Clustering Algorithms}
\textbf{Phase 1: Establishing the Macro-Baseline with K-Means.} We selected K-means as a centroid-based baseline because it provides a simple, interpretable, and computationally efficient partitioning of the feature space into compact groups. This choice is consistent with the flow-based traffic clustering methodology proposed by M\"{u}nz et al. \cite{munz2007traffic}. In the context of MQTT flow analysis, K-means serves as a macro-level baseline by capturing dominant traffic structures and high-volume behavioral groups. The algorithm requires the number of clusters, (K), to be specified in advance, initializes (K) cluster centers, assigns each sample to its nearest centroid, and iteratively updates the centroids until convergence \cite{na2010research}. Since the objective of this study is exploratory rather than supervised classification, we did not constrain (K) to the number of dataset labels. Traffic associated with the same label may exhibit multiple behavioral modes and have 
distinct manifestations of the same attack category. Consequently, restricting (K) to the number of labels could suppress potentially meaningful substructure within the data. To account for this possibility, we evaluated (K=2,\ldots,16). The lower bound of (K=2) corresponds to the minimum number of classes present in the evaluated datasets. The upper bound of (K=16) was selected as a conservative exploratory limit: it substantially exceeds the maximum number of labeled classes in the datasets (6 classes) thereby allowing the discovery of finer-grained behavioral groupings while remaining below a threefold expansion of the maximum label count. This provides sufficient flexibility to identify potential substructure within both benign and malicious traffic without encouraging excessive fragmentation into very small clusters. The chosen range also keeps the model-selection process computationally feasible, as the computational cost of K-means increases with (K) \cite{na2010research}. In our setting, for each value of (K), the algorithm is executed 10 times and the resulting evaluation metrics is averaged to reduce initialization-related variability.

\textbf{Phase 2: Isolating Non-Spherical Anomalies via DBSCAN.} This algorithm was included as a density-based alternative to K-means because it does not require the number of clusters to be specified in advance and can explicitly identify low-density samples as noise. This makes it suitable for exploratory traffic analysis, where benign and malicious MQTT behaviors may not form compact centroid-shaped groups, and where rare or abnormal behaviors may appear either as dense local regions or as isolated outliers. In contrast to centroid-based clustering, DBSCAN defines clusters as dense regions separated by areas of lower density, using two parameters: $\epsilon$, which defines the neighborhood radius around each point, and \textit{MinPts}, which defines the minimum number of samples required for a point to be considered a core point. The cluster is expanded by iteratively adding density-reachable points around the core point. Points that do not satisfy the core point condition and are not density-reachable from any core point are treated as noise \cite{khan2014dbscan}. We therefore evaluated DBSCAN using a grid search over both parameters in order to assess the sensitivity of the clustering results to different density assumptions. The tested $\epsilon$ values were 0.1, 0.3, 0.5, 0.75, 1.0, 1.5, 2.0, and 2.5, covering both strict local-density settings and more permissive neighborhood definitions. The \textit{MinPts} values ranged from 2 to 10, allowing the experiments to capture small dense groups while also testing more restrictive definitions of cluster cores.

\textbf{Phase 3: Hierarchical Extraction of Micro-Clusters via HDBSCAN.} We decided to include HDBSCAN so as to reduce the configuration burden associated with density-based clustering. While DBSCAN requires the system designer to select a global neighborhood radius $\mathrm{Eps}$ and \textit{MinPts} a priori, this choice can be difficult when clusters occur at different densities. HDBSCAN addresses this limitation by extending DBSCAN into a hierarchical density-based method: instead of relying on one fixed global radius, it evaluates the clustering structure across multiple density levels and extracts clusters that remain stable across the hierarchy \cite{stewart2022implementation}. HDBSCAN has been used in the context of malicious and IoT traffic labeling \cite{lo2025hierarchical, garcia2026hybrid} before. This makes it suitable for exploratory MQTT flow analysis, where meaningful traffic groups may exhibit substantially different densities. From a system-design perspective, HDBSCAN shifts the main configuration choice from selecting a distance threshold to selecting the minimum group size that should be considered meaningful. This is easier to interpret operationally, because the designer can reason in terms of the smallest traffic group worth reporting rather than estimating an appropriate geometric radius in a normalized feature space. In this study, we therefore varied $\mathrm{min\_cluster\_size}$ (mcs) from 2 to 60 to evaluate sensitivity to different levels of cluster granularity, ranging from very small behavioral groups to more conservative, stable clusters. Specifically these are the values on which we ran our experiments:
\[mcs \in [2,5,10,15,20,25,30,40,50,60]\] 

\subsubsection{Dimensionality Reduction.}
Finally, to assess the effect of dimensionality reduction, all clustering experiments were conducted both with and without Principal Component Analysis (PCA). PCA was used because the extracted flow features may contain correlated or redundant measurements, especially when multiple tools generate related statistics. 
PCA transforms the original feature space into a smaller set of orthogonal components that preserve the dominant variance structure of the data \cite{abdi2010principal}. In our PCA-based settings, the feature space was reduced to 10, 20, or 30 principal components, depending on the number of valid features available after preprocessing. This allowed us to evaluate whether clustering performance improves when the algorithms operate on a compact representation that retains the main information while reducing noise, redundancy, and dimensionality-related effects. When a dataset did not contain enough valid features to support a given PCA configuration, that configuration was omitted.

\subsection{Evaluation Metrics}
\label{sec:methods:metrics}

We use 2 metrics to evaluate the performance of our algorithms, namely the Silhouette score and the Adjusted Rand Index (ARI). First, using the Silhouette score helps determine which setting of a given algorithm generates the most coherent clusters. 
This metric provides an internal evaluation of cluster quality without requiring class labels. A Silhouette score can fall in the range between -1 and 1; the higher the score, the more coherent the cluster. The lower bound of -1 indicates that the sample is in the wrong cluster, while the upper bound of 1 means that it is in the correct one. The Silhouette score is the mean of all the Silhouette coefficients in the dataset. For each sample, the coefficient is calculated by considering, both the mean intra-cluster distance \textit{a} and the mean nearest-cluster distance \textit{b}. The following equation is used \cite{shahapure2020cluster}: 
\begin{equation}
(b - a)/ max(a,b)
\end{equation}
On the other hand, we use ARI to compare the similarity of the clusters generated by the algorithm -- with a given set of parameters -- to the original labeled classes of the dataset. This method is well established in the literature and has been used previously in the context of malware classification \cite{rashed2021analysis}. ARI's value falls between between -1 and 1. A higher value of ARI means more similarity between the clusters. A value of 1 indicates perfect agreement, values close to 0 indicate agreement similar to random labeling, and negative values may occur when the agreement is worse than expected by chance. The following equation represents ARI \cite{steinley2016variance}:
\begin{equation}
ARI =
\frac{\text{Index} - \text{Expected Value}}
{\text{Maximum Value of Index} - \text{Expected Value}}
\end{equation}

, where \textit{Index} is what is actually observed, \textit{Expected Value} what would be obtained purely by chance and \textit{Maximum Value of Index} is the best possible score.

\section{Results}
\label{sec:res}
In this section, we highlight the results of our clustering experiments on each of the datasets of the generated flows by our 3 tools. Tables \ref{tab:kmeans_res}, \ref{tab:dbscan_res} and \ref{tab:hdbscan_res} show the results of our experiments. 

\subsection{K-means}
\label{sec:res:kmeans}

To avoid any random initialization in our results, for each K-value (2-16), we ran our script 10 times and averaged the output. We observe that most of the top Silhouette and ARI values were obtained with Zeek with the PCA10 setting. The exceptions in which Zeek PCA 10 was not the top performer for the Silhouette score were: a) NTLFlowLyzer for the GothX (also with PCA10) and b) Zeek with PCA20 for the MQTTset. In the case of ARI, the only exception was in the case of MQTT-IoT-IDS2020, in which Zeek with the original features was the best performer (no dimensionality reduction). We note that in the cases of Artemis and MQTT-IoT-IDS2020 the best K for Silhouette and ARI was 2, similar to the number of original classes in these datasets. For GothX, which has 6 classes, the best K for Silhouette was at K=12 and for ARI was at K=8. For MQTTset, which has 4 original classes, the best K for Silhouette was at K=2, while the best for ARI was at K=9. Another observation is that Artemis has the lowest scores for Silhouette and ARI among the other datasets. Table \ref{tab:kmeans_res} demonstrates the results for the K-means algorithm experiments.

We note that, we did not run PCA 30 for MQTTset because the number of columns with non-unique values were 28 (less than the needed 30 columns) to run the algorithm with this setting.  

\subsection{HDBSCAN}
\label{sec:res:hdbscan}
We observe that most of the top ARI values were obtained with Zeek utilizing reduced dimensions (2 with PCA10 and 1 with PCA30). The only exception for ARI was the Artemis dataset, in which Argus was the best performer using the original features (no dimensionality reduction). For the Silhouette score, the top performers were evenly split: Zeek achieved the highest scores for GothX (PCA10) and MQTT-IoT-IDS2020 (PCA20), while Argus achieved the highest scores for Artemis (PCA20) and the MQTTset, reaching a perfect score of 1.0 with the original features. We note that the optimal MCS parameter for the best Silhouette score was 2 for both Artemis and MQTTset, while it was much higher in the cases of GothX (30) and MQTT-IoT-IDS2020 (50). 
For ARI, however, the optimal MCS varied widely, ranging from 5 for MQTTset to as high as 60 for MQTT-IoT-IDS2020. An additional note, similar to the K-means experiments, the best Silhouette and ARI values for Artemis is the least among all the datasets. The ARI score in this case was very low (0.02) when compared to the next lowest value 0.52 (MQTTset). Table \ref{tab:hdbscan_res} demonstrates the results for the HDBSCAN algorithm experiments. 

\subsection{DBSCAN}
\label{sec:res:dbscan}
Unlike the previous experiments, DBSCAN best performing tools varied 
Similarly, for the DBSCAN experiments, we observe that most of the top ARI values were obtained with Zeek using reduced dimensions (PCA10 and PCA30), though it tied with Argus (with no dimensionality reduction) on the MQTTset dataset. The exception for ARI was again the Artemis dataset, where Argus was the best performer utilizing the original features. For the Silhouette score, top performers were more varied: NTLFlowLyzer was the best performer for GothX (PCA20) and MQTTset (PCA10), while Argus was the top performer for Artemis (PCA30). Zeek (PCA30) was the best performer in the case of MQTT-IoT-IDS2020. Unlike K-means, we note that the optimal number of clusters generated by DBSCAN for the best ARI drastically exceeded the actual number of classes across all datasets. For Artemis and MQTT-IoT-IDS2020 (both containing 2 classes), the best ARI was achieved with 21 and 34 clusters, respectively. For GothX (6 classes) and MQTTset (4 classes), the highest ARI produced 55 and 15 clusters for Zeek, respectively. We also observe that the configurations yielding the highest ARI consistently isolated a very low percentage of noise, remaining mostly under 1\%. Finally, similar to the previous experiments, Artemis has the worst best performance among the other datasets for both ARI and Silhouette. It's best ARI score is 0.13, while the next least score is 0.91. Table \ref{tab:dbscan_res} demonstrates the results for the DBSCAN algorithm experiments.  

\begin{table}[ht]
\centering
\small
\setlength{\tabcolsep}{5pt}
\begin{tabular}{llllr|llrr}
\multicolumn{2}{c}{} & \multicolumn{3}{c|}{\textbf{SILHOUETTE}} & \multicolumn{3}{c}{\textbf{ARI}} & \multicolumn{1}{c}{} \\ \cline{3-8}
\multicolumn{1}{c}{\textbf{Dataset}} & \multicolumn{1}{c}{\textbf{Tool}} & \textbf{val} & \textbf{dim} & \textbf{K} & \textbf{val} & \textbf{dim} & \textbf{K} & \multicolumn{1}{c}{\multirow{-2}{*}{\textbf{clas}}} \\ \Xhline{2pt}

 & argus & 0.58 & pca\_10 & 14 & 0.07 & pca\_10 & 16 &  \\ \cline{2-8}
 Artemis & Lyzer & 0.50 & pca\_10 & 2 & 0.13 & pca\_10 & 2 & 2 \\ \cline{2-8}
 & Zeek & \textbf{0.63} & pca\_10 & 2 & \textbf{0.27} & pca\_10 & 2 &  \\ \Xhline{1.5pt}

 & argus & 0.83 & pca\_10 & 2 & 0.96 & original & 9 &   \\ \cline{2-8}
GothX & Lyzer &\textbf{0.96} & pca\_10 & 12 & 0.22 & pca\_10 & 9 & 6 \\ \cline{2-8}
 & Zeek & 0.94 & pca\_10 & 2 & \textbf{0.97} & pca\_10 & 8 &  \\ \Xhline{1.5pt}

\multirow{3}{*}{\begin{tabular}[c]{@{}l@{}}MQTT-IoT-\\ IDS2020\end{tabular}} & argus & 0.56 & pca\_10 & 3 & 0.34 & pca\_10 & 4 &  \\ \cline{2-8} 
 & Lyzer & 0.69 & pca\_10 & 4 & 0.38 & pca\_10 & 2 & 2 \\ \cline{2-8} 
 & Zeek & \textbf{0.84} & pca\_10 & 2 & \textbf{0.96} & original & 2 &  \\ \Xhline{1.5pt}

\multirow{3}{*}{\begin{tabular}[c]{@{}l@{}}MQTTset\\ \end{tabular}} & argus & 0.83 & pca\_10 & 2 & 0.78 & original & 15 &  \\ \cline{2-8} 
 & Lyzer & 0.79 & pca\_10 & 14 & 0.41 & pca\_30 & 13 & 4 \\ \cline{2-8} 
 & Zeek & \textbf{0.88} & pca\_20 & 2 & \textbf{0.82} & pca\_10 & 9 &  \\ \Xhline{2pt}

\end{tabular}
\caption{K-means's top Silhouette and ARI values for each tool along with the corresponding number of dimensions, the K-value at which the best score occurs. \textit{clas} = \textbf{number of classes in dataset}.}
\label{tab:kmeans_res}
\end{table}

\begin{table}[ht]
\centering
\small
\setlength{\tabcolsep}{5pt}
\begin{tabular}{llllr|llrr}
\multicolumn{2}{c}{} & \multicolumn{3}{c|}{\textbf{SILOUETTE}} & \multicolumn{3}{c}{\textbf{ARI}} & \multicolumn{1}{c}{} \\ \cline{3-8}
\multicolumn{1}{c}{\textbf{set}} & \multicolumn{1}{c}{\textbf{Tool}} & \textbf{val} & \textbf{dim} & \textbf{mcs} & \textbf{val} & \textbf{dim} & \textbf{mcs} & \multicolumn{1}{c}{\multirow{-2}{*}{\textbf{clas}}} \\ 
\Xhline{1.5pt}
 & argus & \textbf{0.47} & p20 & 2 & \textbf{0.02} & orig & 20 &  \\ \cline{2-8} 
Artemis & Lyzer & 0.35 & p10 & 10 & 0.0001 & orig & 2 & 2 \\ \cline{2-8} 
 & Zeekk & 0.39 & p10 & 5 & 0.0001 & orig & 2 &  \\ \Xhline{1.5pt}

 & argus & 0.77 & p20 & 60 & 0.95 & orig & 30 &  \\ \cline{2-8} 
GothX & Lyzer & 0.50 & orig & 60 & 0.13 & p20 & 15 & 6 \\ \cline{2-8} 
 & Zeek & \textbf{0.93} & p10 & 30 & \textbf{0.97} & p10 & 25 &  \\ \Xhline{1.5pt}

\multirow{3}{*}{\begin{tabular}[c]{@{}l@{}}MQTT-IoT-\\ IDS2020\end{tabular}} & argus & 0.46 & p10 & 60 & 0.10 & p10 & 60 & \\ \cline{2-8} 
 & Lyzer & 0.49 & orig & 2 & 0.20 & orig & 40 & 2 \\ \cline{2-8} 
 & Zeek & \textbf{0.57} & p20 & 50 & \textbf{0.85} & p10 & 60 & \\ \Xhline{1.5pt}

\multirow{3}{*}{\begin{tabular}[c]{@{}l@{}}MQTTset\\ \end{tabular}} & argus & \textbf{1.00} & orig & 2 & 0.49 & orig & 50 &  \\ \cline{2-8} 
 & Lyzer & 0.56 & orig & 40 & 0.23 & p20 & 20 & 4 \\ \cline{2-8} 
 & Zeek & 0.41 & p30 & 10 & \textbf{0.52} & p30 & 5 &  \\ \Xhline{1.5pt}

\end{tabular}
\caption{HDBSCAN's top Silhouette and ARI values for each tool along with the number of dimensions, the parameters and the percentage of samples that are considered to be noise. 
\textit{dim} = \textbf{dimensions} with the values p10=PCA10,etc. and orig=original dimensions without reduction. \textit{mcs} = \textbf{minimum cluster size}, \textit{clas} = \textbf{number of classes in dataset}.}
\label{tab:hdbscan_res}
\end{table}

\begin{table}[ht]
\centering
\small 
\setlength{\tabcolsep}{3pt} 
\begin{tabular}{llllrrrr|llrrrrr}

\multicolumn{1}{c}{} & \multicolumn{1}{c}{} & \multicolumn{6}{c|}{SILHOUETTE} & \multicolumn{6}{c}{ARI} & \multicolumn{1}{c}{} \\ \cline{3-14}
\multicolumn{1}{c}{\multirow{-2}{*}{\rotatebox{90}{set}}} & \multicolumn{1}{c}{\multirow{-2}{*}{\rotatebox{90}{tool}}} & val & dim & eps & m.pt & clus & ns\% & val & dim & eps & m.pt & clus & ns\% & \multicolumn{1}{c}{\multirow{-2}{*}{\rotatebox{90}{{clas}}}} \\ \Xhline{1.5pt}
\multirow{3}{*}{\rotatebox{90}{Artemis}} & argus & \textbf{0.55} & p30 & 0.5 & 7 & 68 & 5.97 & \textbf{0.13} & orig & 2.5 & 4 & 21 & 0.32 &  \\ \cline{2-14} 
 & Lyzer & 0.43 & p10 & 2.5 & 6 & 5 & 3.64 & 0.01 & p10 & 2.5 & 2 & 12 & 1.45 & 2 \\ \cline{2-14} 
 & Zeek & 0.46 & p10 & 2 & 4 & 4 & 3.71 & 0.003 & p10 & 2.5 & 2 & 11 & 1.96 &  \\ \Xhline{1.5pt}
 \multirow{3}{*}{\rotatebox{90}{GothX}}& argus & 0.78 & orig & 2.5 & 6 & 32 & 0.60 & 0.96 & p10 & 1.5 & 2 & 60 & 0.29 &  \\ \cline{2-14} 
 & Lyzer & \textbf{0.94} & p20 & 1.5 & 6 & 20 & 0.17 & 0.22 & p20 & 1.5 & 2 & 41 & 0.07 & 6 \\ \cline{2-14} 
 & Zeek & 0.93 & p10 & 2 & 5 & 4 & 0.15 & \textbf{0.97} & p30 & 1.5 & 2 & 55 & 0.24 &  \\ \Xhline{1.5pt}
\multirow{3}{*}{\rotatebox{90}{MQTT-IoT-IDS20}} & \vspacing{15pt}{10pt} argus & 0.47 & p10 & 1 & 9 & 21 & 1.02 & 0.10 & orig & 1.5 & 2 & 77 & 0.35 &  \\ \cline{2-14} 
& \vspacing{15pt}{10pt} Lyzer & 0.52 & p10 & 2 & 10 & 8 & 0.15 & 0.24 & p10 & 2.5 & 3 & 10 & 0.02 & 2 \\ \cline{2-14} 
 &\vspacing{15pt}{10pt} Zeek & \textbf{0.58} & p30 & 2.5 & 10 & 8 & 1.45 & \textbf{0.91} & p10 & 2.5 & 3 & 34 & 0.65 &  \\ \Xhline{1.5pt}
 \multirow{3}{*}{\rotatebox{90}{MQTTset}} & argus & 0.69 & p10 & 2 & 3 & 19 & 0.55 & \textbf{0.91} & orig & 1 & 5 & 27 & 2.64 &  \\ \cline{2-14} 
 & Lyzer & \textbf{0.78} & p10 & 2 & 10 & 15 & 1.52 & 0.44 & p20 & 2.5 & 9 & 17 & 1.99 & 4 \\ \cline{2-14} 
& Zeek & 0.64 & p10 & 2.5 & 4 & 7 & 1.30 & \textbf{0.91} & p10 & 1.5 & 2 & 15 & 0.95 &  \\ 
 \Xhline{1.5pt}
\end{tabular}
\caption{DBSCAN's top Silhouette and ARI values for each tool along with the number of dimensions, the parameters and the percentage of samples that are considered to be noise. \textit{dim} = \textbf{dimensions} with the values p10=PCA10,etc. and orig=original dimensions without reduction. \textit{eps} = \textbf{epsilon}, \textit{mnPt} = \textbf{minimum Points}, \textit{clus} = \textbf{number of clusters}, \textit{ns\%} = \textbf{percentage of noise} and \textit{clas} = \textbf{number of classes in dataset}.}
\label{tab:dbscan_res}
\end{table}

\subsection{Efficiency and Scalability}
\label{sec:res:perf}
The experimental phase was conducted on a 64-bit Linux server powered by a 32-core AMD EPYC 9354P processor (3.25 GHz) with 315 GB of RAM. To manage computational resources and ensure memory stability on the 32-core node, different parallelization strategies were employed: the K-means experiments utilized full multi-core acceleration, while DBSCAN was restricted to 2 concurrent jobs to prevent memory exhaustion, and HDBSCAN was executed in a single-threaded configuration. Opting for using all the cores in the K-means experiments was due to the delay caused by single processing. It is very well known that K-means is not the best algorithm when it comes to scalability; in the worst case scenario, it can be exponential \cite{vattani2009k}. On the other hand, the decision to limit DBSCAN was based on several preliminary experiments. For HDBSCAN, it was totally functional in the default configuration. Thus, we chose to leave it as is.

\begin{figure}
\centering
\begin{subfigure}{0.59\textwidth}
  \centering
  \includegraphics[width=\textwidth]{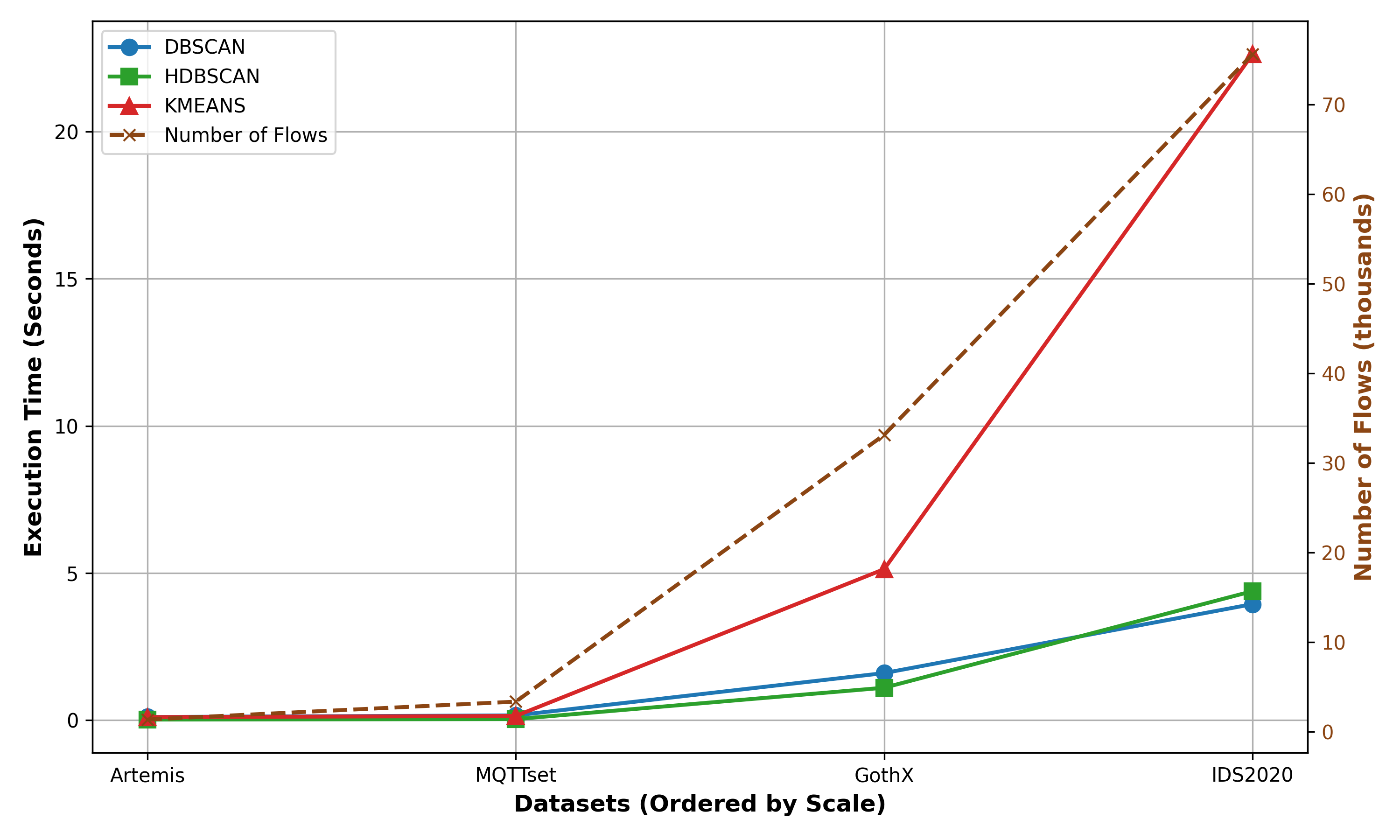}
  \vspace{-16pt}
  \caption{Zeek}
  \label{fig:timeZeek}
 \end{subfigure}%
 
\begin{subfigure}{0.59\textwidth}
  \centering
  \includegraphics[width=\textwidth]{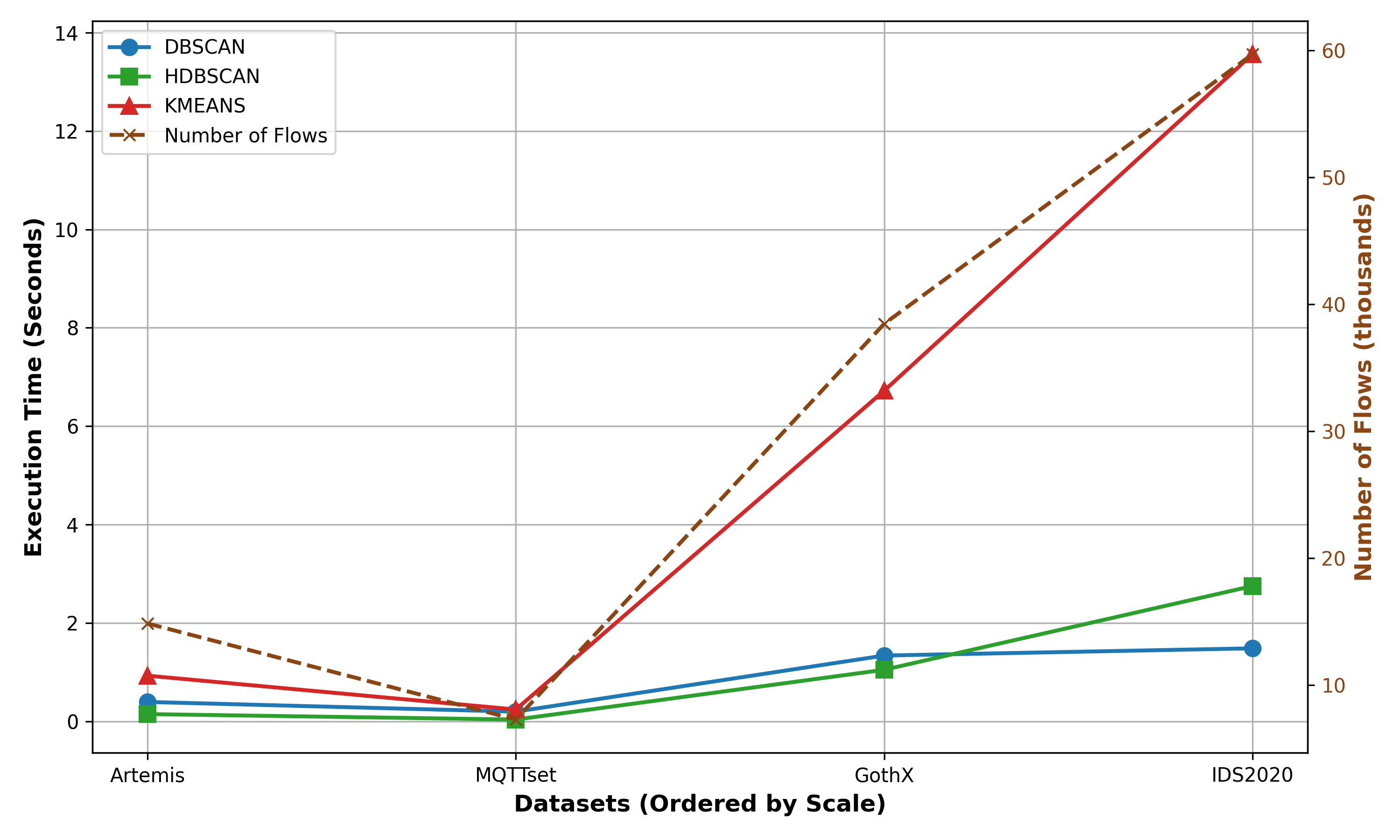}
  \vspace{-16pt}
  \caption{Argus}
  \label{fig:timeArgus}
\end{subfigure}
\begin{subfigure}{0.59\textwidth}
  \centering
  \includegraphics[width=\textwidth]{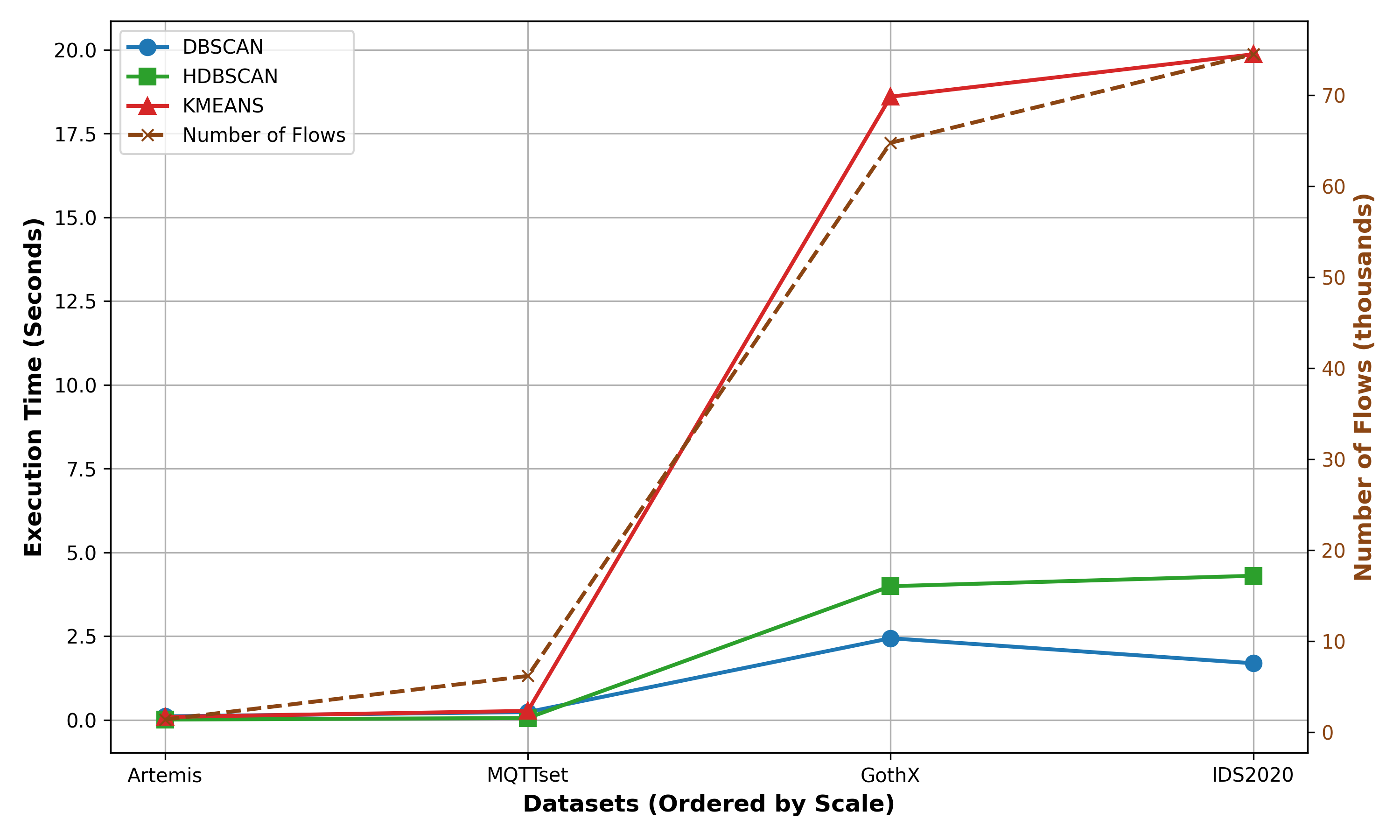}
  \vspace{-16pt}
  \caption{NTLFlowLyzer}
  \label{fig:timeLyzer}
\end{subfigure}
\vspace{-8pt}
\caption{Execution Times for the algorithms per tool vs the number of flows}
\label{fig:exectimes}
\end{figure}

The computational efficiency for Zeek-derived flows is illustrated in Fig. \ref{fig:timeZeek}. As the data scales from around 1.4K flows (Artemis) to over 75K flows (MQTT-IoT-IDS2020), we observe an unusual trend where K-means demonstrates as the most time-intensive algorithm. In contrast, DBSCAN and HDBSCAN maintain lower execution time. Despite showing increasing execution time as the number of flows grows, the curves of DBSCAN and HDBSCAN rise less steeply than the K-means curve in Figure \ref{fig:exectimes}. 
The same phenomenon occurs in the cases of the Argus and NTLFlowLyzer flows (see Figures \ref{fig:timeArgus} and \ref{fig:timeLyzer}). While K-means execution time follows the number of flows curve, both DBSCAN and HDBSCAN have a much slower increase compared to that of the number of flows. Besides, we note that for the Zeek and Argus flows, HDBSCAN's execution time is very similar (slightly less) than DBSCAN for the datasets Artemis, MQTTset and GothX, while it slightly bigger for MQTT-IoT-IDS2020 (the largest dataset). Yet, for NTLFlowLyzer's flows, the execution time is less for the Artemis and MQTTset sets, and is more for both GothX and MQTT-IoT-IDS2020. Overall, we observe that DBSCAN's curves for flows of the 3 tools have a slower increase in execution time than HDBSCAN.

To better compare the impact of algorithms vs. tools on the execution time, we calculated the average throughput (flows per second) for each combination of tool+algorithm using the following equation:

\begin{equation}
    \overline{\eta}_{T, A} = \frac{1}{n} \sum_{i=1}^{n} \frac{N_{i,T}}{t_{i,T,A}}
    \label{eq:avg_efficiency}
\end{equation}

where:
\begin{itemize}
    \item $n$ is the total number of datasets ($n=4$),
    \item $i$ denotes the specific dataset index,
    \item $T$ represents the flow extraction tool (e.g., Zeek, Argus, or NTLFlowLyzer),
    \item $A$ represents the clustering algorithm (DBSCAN, HDBSCAN, or K-means),
    \item $N_{i,T}$ is the number of flows generated by tool $T$ for dataset $i$, and
    \item $t_{i,T,A}$ is the corresponding execution time for algorithm $A$ when processing the data from tool $T$ on dataset $i$.
\end{itemize}

\begin{table}[htbp]
    \centering
    \begin{tabular}{lcccc}
        \hline
        \textbf{Algorithm ($A$)} & \textbf{All} & \textbf{Argus} & \textbf{Flowlyzer} & \textbf{Zeek} \\
        \hline
        $\overline{\eta}_{T, \text{DBSCAN}}$  & 27.34 & 35.81 & 27.34 & 18.88 \\
        $\overline{\eta}_{T, \text{HDBSCAN}}$ & 71.44 & 88.79 & 60.96 & 64.55 \\
        $\overline{\eta}_{T, \text{K-means}}$ & 12.54 & 14.07 & 11.32 & 12.23 \\
        \hline
        \textbf{Mean Tool Efficiency ($\overline{\eta}_{T}$)} & \textbf{37.11} & \textbf{46.22} & \textbf{33.20} & \textbf{31.89} \\
        \hline
    \end{tabular}
    \caption{Average Computational Throughput ($\overline{\eta}$) measured in flows per second across different tools. Higher values indicate greater processing efficiency.}
    \label{tab:thruput_measures}
\end{table}

While Figure \ref{fig:exectimes} summarizes the execution time as indicated earlier, Table \ref{tab:thruput_measures}  averages throughput across clustering algorithms and flow-generation tools. K-means showed the highest execution times and the lowest throughput across the evaluated datasets, followed by DBSCAN and finally HDBSCAN.

\subsubsection{Correlations between Flows Count vs. Algorithm Execution Time}
To quantify the scalability of our approach, we calculated the correlation coefficients between the number of flows and algorithm execution time (see Table \ref{tab:correlations}). The results demonstrate extremely high positive correlation values for all tools and algorithms \textit{(r>0.89)}. The results show a near-perfect positive correlation \textit{(r>0.97)} for both K-means and HDBSCAN across all flow extraction tools, confirming that execution time scales predictably with dataset volume. DBSCAN exhibits a slightly lower correlation in the aggregate 'All' category \textit{(r=0.895)}, suggesting its performance is slightly less sensitive to the flow count compared to the other algorithms. Overall, the consistently high correlation values are a clear evidence that the flow volume is the primary determinant of computational overhead in the analysis pipeline.

\begin{table}[htbp]
    \centering
    \begin{tabular}{lcccc}
        \hline
        \textbf{Algorithm} & \textbf{All} & \textbf{Argus} & \textbf{FlowLyzer} & \textbf{Zeek} \\
        \hline
        DBSCAN  & 0.895 & 0.961 & 0.930 & 0.999 \\
        HDBSCAN & 0.974 & 0.974 & 0.998 & 0.983 \\
        K-Means  & 0.980 & 0.992 & 0.997 & 0.978 \\
        \hline
    \end{tabular}
    \caption{Correlation between \textit{No. of Flows} and \textit{Algorithm Execution Time} for each of the tools.}
    \label{tab:correlations}
\end{table}

\section{Discussion}
\label{sec:discussion}

Although supervised ML approaches could have produced higher accuracy and precision values, such a setting is less aligned with the objective of this study. Our goal is not to build an optimized detector under fully labeled conditions, but to characterize MQTT traffic from a realistic operational perspective, where network administrators often face heterogeneous traffic with incomplete or uncertain labels. For this reason, we adopt an unsupervised learning approach and evaluate the resulting structure using ARI and Silhouette score, which respectively measure agreement with available labels and internal cluster compactness/separation.

We use traffic flows rather than only raw packets because flows provide a connection-level view of network behavior. While raw packets describe individual protocol events, flow records summarize communication over time through features such as duration, packet and byte counts, directionality, rates, timing statistics, and endpoint interaction patterns. This aggregation allows us to analyze broader MQTT traffic dynamics across datasets and flow-generation tools, rather than treating packets as isolated observations. In this sense, our study approaches MQTT traffic analysis from a real-world characterization perspective rather than a “perfect-results” supervised classification perspective.

With this philosophy in mind, our experiments yielded several interesting findings. First, while the number of flows for each pcap varied by the tool as shown in Table \ref{tab:flows}, we found that the PDF of \(\frac{\textit{no.\_flows\_Zeek}}{\textit{no.\_flows\_FlowLyzer}}\)
\vspace{0.5em} 
is highly concentrated around the 0.5 area. On the other hand, the \textit{Zeek\textdiv Argus} and \textit{Argus\textdiv FlowLyzer} were more scattered (see Figure \ref{fig:pdfsFlows}). This indicates a relatively stable proportional relationship between the number of flows generated by Zeek and NTLFlowLyzer. However, this relationship does not hold uniformly across all captures. For example, Table \ref{tab:flows} shows that in some cases this relationship does not hold (e.g. GothX's Slowite) where the 2 tools yield the same number of flows or even Zeek produces more flows than NTLFlowLyzer (e.g. MQTTset's malariaDoS). The same applies to the relationship between Argus and Zeek, and between Argus and NTLFlowLyzer. For the latter relationships (both involving Argus), it is observable that the number of flows vary drastically. For instance, Zeek and Argus both have the same number of flows for GothX's normal pcap, while the number of flows of Argus is more than 1.8K times of Zeek's for the Artemis benign's pcap. 

With regards to the experiments, as highlighted in Section \ref{sec:res}, the best performing tool is notably Zeek at the PCA 10 setting. The best results are obtained with K-means. Moreover, generally, the number of clusters (K) is much less than those of the best scores of DBSCAN. However, K-means is designed to always have the samples belonging to a cluster. This is different from DBSCAN which considers the samples that it fails to include in a cluster to be noise. Similarly, HDBSCAN requires a single parameter (\textit{min.\_cluster\_size}). Despite the algorithm being required to assign each sample to a cluster, our experimental design has allowed for testing the algorithm on a wide range of that parameter (2 ... 60). The difference is reflected in the number of clusters that generate the top scores; for K-means it is mostly below 10 clusters, while for HDBSCAN it is mostly over 20 clusters.

To draw a full picture of the reason behind these variations with regards to the number of flows and very likely with the performance of the tools, it is essential to take into account the modi operandi of each of the tools described in Sections \ref{sec:tools:argus}, \ref{sec:tools:lyzer} and \ref{sec:tools:zeek}. 
These modi operandi help us determine several factors that are likely to lead to differences in both the flow-generation and the clustering performance: 
\begin{itemize}
\item While Argus has a record interval of 60 seconds, essentially splitting flows that surpass that time windows, this differs completely from how both Zeek and NTLFlowLyzer operate. NTLFlowLyzer have an activity timeout that is much larger (5K seconds) and splits a flow when passes the 120K seconds mark. On the other hand, Zeek has an activity timeout of 300 seconds and never splits an ongoing flow. 
That consequently means that real flows that last more than 60 seconds are not really reflected as such when argus's default setting is used. For NTLFlowLyzer, in the case of flows > 120K seconds, a split occurs. Yet, the long duration of 120K seconds (33 hours and 20 minutes) makes the probability of splitting extremely low. In this context, Zeek stands out for the lack of splitting of flows. While more analysis is needed, we deduce that how flows are split by each of the 3 tools is reflected in the concentration of the ratio between the number of flows for
NTLFlowLyzer and Zeek around the center of Figure \ref{fig:pdfsFlows} unlike the ratios that involve argus. 

\item On a deeper level, for short-lived TCP connections, the last packet in the record may correspond to the actual connection termination packet in the specific case of argus. In contrast, for long-lived TCP connections, the last packet in a given record may correspond only to the last packet observed before Argus emitted a periodic status record, rather than to the end of the underlying TCP connection. This influence consequently has drastic effect on how the used algorithm views the flows. The differences in the structure of the flows essentially affects how the algorithm views the samples in the dataset and determines the weights of features to decide on samples assignation for clusters.

\item Another factor that affects the performance is related to the conditions upon how each of the tools decides the start and end of a flow or discards packets (e.g. Zeek discards trailing non-productive packets after one direction has closed the connection). In the particular example of Zeek removal of tailing non-productive packets like final ACK, the impact would not be that big. However, other conditions would definitely have a bigger influence on how start/end of flows are decided. This consequently affects how algorithms decides on sample membership in one cluster or another.
\end{itemize}

Finally, with regards to the time efficiency,  Table 6 suggests that, within this experimental pipeline, the choice of clustering algorithm affected throughput more strongly than the choice of flow-generation tool. However, this pattern may partly reflect the different parallelization strategies and search spaces used for each of the 3 clustering algorithms
. We note that DBSCAN and HDBSCAN appear to scale more favorably than K-means under our implementation and parameter settings. Yet, this observation should be interpreted cautiously because the algorithms were executed under different parallelization settings, and K-means was also evaluated over multiple values of (K) with repeated runs per configuration.

\subsection{Limitations}
\label{sec:limit}
This study has several limitations. First, it focuses only on unencrypted MQTT traffic over port 1883. As noted in Section \ref{sec:method:scope}, we do not analyze encrypted MQTT traffic as it involves additional considerations related to the impact of the use TLS. Thus, MQTTS traffic analysis is left for future work. Additionally, the analysis is limited to 4 publicly available MQTT datasets and three flow-generation tools. The choice of the datasets and the tools is more representative than exhaustive. The work of \cite{haddadi2014benchmarking} refers to other similar tools, which we shall test in future work. Also, the tools were used with default flow-construction settings because the aim of this study was to compare representative out-of-the-box behavior before introducing tool tuning as an additional experimental factor. Consequently, alternative inactivity timeouts, periodic reporting intervals, or maximum-duration limits may produce different flow populations and clustering outcomes. A systematic sensitivity analysis of these parameters is therefore left for future work. 
Moreover, ARI relies on the original dataset labels, which may not always map perfectly to flow-level behavior. Furthermore, for MQTT-IoT-IDS2020, the proportional subset was selected by retaining the first records of each class in the exported CSV files. This choice was intended to preserve the temporal ordering of the generated flows. However, because each flow-generation tool applies its own flow construction, timeout, and export mechanisms, the retained records are not guaranteed to correspond to the same underlying packet intervals across tools. As a result, a flow included in the subset generated by one tool may not have a direct counterpart in the subset generated by another tool. Finally, some packet captures were excluded when they could not be processed by all three tools, which may have likely limited the coverage of our analysis.

\section{Conclusions}
\label{sec:conc}
In this work, we presented a cross-dataset, tool-aware characterization of MQTT traffic using both benign and malicious traffic from four publicly available MQTT datasets. Rather than proposing another supervised intrusion detector, our objective was to examine how MQTT traffic is structured when raw packet captures are transformed into flow records, and how this structure changes across datasets, flow-generation tools, clustering algorithms, and feature-space configurations.

To this end, we converted the selected packet captures into traffic flows using three flow-generation tools: Zeek-flowmeter, Argus, and NTLFlowLyzer. These tools collectively generated more than 380K flow records. However, the number and structure of the generated flows varied substantially from one tool to another, confirms that flow extraction is not a neutral preprocessing step but rather relies on the internal rules and mechanisms of each flow-generation tool.

We analyzed the resulting flow representations using K-means, DBSCAN, and HDBSCAN, with and without PCA-based dimensionality reduction. The results show that clustering behavior is affected not only by the dataset and the algorithm, but also by the flow-generation tool. Zeek-flowmeter achieved the strongest performance in most experiments, particularly when combined with PCA-10 setting. Yet, no single tool consistently dominated across all datasets, algorithms, and evaluation metrics, thus confirming that all the factors of dataset, algorithm and flow-generation tool influence performance. 

Overall, the study shows that flow-level MQTT traffic analysis should account for the full experimental pipeline: the dataset, the flow-generation tool, the preprocessing choices, the clustering algorithm, and the evaluation metric. Our future work will extend this analysis to encrypted MQTT traffic, additional flow-generation tools, and alternative tool configurations. We also plan to conduct deeper feature-level analysis to identify which flow attributes most strongly influence clustering behavior, and to further examine how benign and attack traffic generation methods affect the observed separability of MQTT traffic.



\section*{Acknowledgements}
This work is supported by grant DISCOVERY (PID2023-148716OB-C33) funded by MCIU/AEI/10.13039/ 501100011033, and by the Comunidad de Madrid (Spain) under the project RAMONES-CM (TEC2024-COM504), co-financed by European Structural Funds (ESF and FEDER).
This work is part of the I-Shaper Project (C114/23), a collaboration agreement between Instituto Nacional de Ciberseguridad (INCIBE) and Universidad Carlos III de Madrid. This initiative is being carried out within the framework of the Recovery, Transformation and Resilience Plan funds, funded by the European Union (Next Generation). Finally,
funding for APC: Universidad Carlos III de Madrid  (Agreement CRUE-Madroño 2026).

\printcredits


\bibliographystyle{unsrtnat}

\bibliography{cas-refs}

@inproceedings{shahapure2020cluster,
  title={Cluster quality analysis using silhouette score},
  author={Shahapure, Ketan Rajshekhar and Nicholas, Charles},
  booktitle={2020 IEEE 7th international conference on data science and advanced analytics (DSAA)},
  pages={747--748},
  year={2020},
  organization={IEEE}
}

@article{ring2019survey,
  title={A survey of network-based intrusion detection data sets},
  author={Ring, Markus and Wunderlich, Sarah and Scheuring, Deniz and Landes, Dieter and Hotho, Andreas},
  journal={Computers \& security},
  volume={86},
  pages={147--167},
  year={2019},
  publisher={Elsevier}
}

@inproceedings{vattani2009k,
  title={K-means requires exponentially many iterations even in the plane},
  author={Vattani, Andrea},
  booktitle={Proceedings of the twenty-fifth annual symposium on Computational geometry},
  pages={324--332},
  year={2009}
}

@inproceedings{erman2006traffic,
  title={Traffic classification using clustering algorithms},
  author={Erman, Jeffrey and Arlitt, Martin and Mahanti, Anirban},
  booktitle={Proceedings of the 2006 SIGCOMM workshop on Mining network data},
  pages={281--286},
  year={2006}
}

@article{haddadi2014benchmarking,
  title={Benchmarking the effect of flow exporters and protocol filters on botnet traffic classification},
  author={Haddadi, Fariba and Zincir-Heywood, A Nur},
  journal={IEEE Systems journal},
  volume={10},
  number={4},
  pages={1390--1401},
  year={2014},
  publisher={IEEE}
}

@article{boppana2023gan,
  title={GAN-AE: An unsupervised intrusion detection system for MQTT networks},
  author={Boppana, Tej Kiran and Bagade, Priyanka},
  journal={Engineering Applications of Artificial Intelligence},
  volume={119},
  pages={105805},
  year={2023},
  publisher={Elsevier}
}

@inproceedings{ranjalahy2025hybrid,
  title={Hybrid Anomaly Detection in MQTT Protocol Using VAE and KNN},
  author={Ranjalahy, Elisa and Rigaux, Martin and Salem, Osman and Mehaoua, Ahmed},
  booktitle={2025 IEEE International Conference on E-health Networking, Application \& Services (Healthcom)},
  pages={1--6},
  year={2025},
  organization={IEEE}
}

@article{abdi2010principal,
  title={Principal component analysis},
  author={Abdi, Herv{\'e} and Williams, Lynne J},
  journal={Wiley interdisciplinary reviews: computational statistics},
  volume={2},
  number={4},
  pages={433--459},
  year={2010},
  publisher={Wiley Online Library}
}

@article{stewart2022implementation,
  title={An implementation of the HDBSCAN* clustering algorithm},
  author={Stewart, Geoffrey and Al-Khassaweneh, Mahmood},
  journal={Applied Sciences},
  volume={12},
  number={5},
  pages={2405},
  year={2022},
  publisher={MDPI}
}

@inproceedings{khan2014dbscan,
  title={DBSCAN: Past, present and future},
  author={Khan, Kamran and Rehman, Saif Ur and Aziz, Kamran and Fong, Simon and Sarasvady, Sababady},
  booktitle={The fifth international conference on the applications of digital information and web technologies (ICADIWT 2014)},
  pages={232--238},
  year={2014},
  organization={IEEE}
}

@inproceedings{na2010research,
  title={Research on k-means clustering algorithm: An improved k-means clustering algorithm},
  author={Na, Shi and Xumin, Liu and Yong, Guan},
  booktitle={2010 Third International Symposium on intelligent information technology and security informatics},
  pages={63--67},
  year={2010},
  organization={Ieee}
}

@article{khan2021deep,
  title={A deep learning-based intrusion detection system for MQTT enabled IoT},
  author={Khan, Muhammad Almas and Khan, Muazzam A and Jan, Sana Ullah and Ahmad, Jawad and Jamal, Sajjad Shaukat and Shah, Awais Aziz and Pitropakis, Nikolaos and Buchanan, William J},
  journal={Sensors},
  volume={21},
  number={21},
  pages={7016},
  year={2021},
  publisher={MDPI}
}

@article{syed2020denial,
  title={Denial of service attack detection through machine learning for the IoT},
  author={Syed, Naeem Firdous and Baig, Zubair and Ibrahim, Ahmed and Valli, Craig},
  journal={Journal of Information and Telecommunication},
  volume={4},
  number={4},
  pages={482--503},
  year={2020},
  publisher={Taylor \& Francis}
}

@article{dikii2021attacks,
  title={DoS attacks detection in MQTT networks},
  author={Dikii, Dmitrii and Arustamov, Sergey and Grishentsev, Aleksey},
  journal={Indonesian Journal of Electrical Engineering and Computer Science},
  volume={21},
  number={1},
  pages={601--608},
  year={2021},
  publisher={Institute of Advanced Engineering and Science}
}

@article{ap2019secure,
  title={Secure-MQTT: an efficient fuzzy logic-based approach to detect DoS attack in MQTT protocol for internet of things},
  author={Ap, Haripriya and K, Kulothungan},
  journal={EURASIP Journal on Wireless Communications and Networking},
  volume={2019},
  number={1},
  pages={90},
  year={2019},
  publisher={Springer}
}

@article{hintaw2019brief,
  title={A Brief Review on MQTT's Security Issues within the Internet of Things (IoT).},
  author={Hintaw, Ahmed J and Manickam, Selvakumar and Karuppayah, Shankar and Aboalmaaly, Mohammed Faiz},
  journal={J. Commun.},
  volume={14},
  number={6},
  pages={463--469},
  year={2019}
}

@article{shafi2025ntlflowlyzer,
  title={NTLFlowLyzer: Towards generating an intrusion detection dataset and intruders behavior profiling through network and transport layers traffic analysis and pattern extraction},
  author={Shafi, MohammadMoein and Lashkari, Arash Habibi and Roudsari, Arousha Haghighian},
  journal={Computers \& Security},
  volume={148},
  pages={104160},
  year={2025},
  publisher={Elsevier}
}

@article{vaccari2020mqttset,
  title={MQTTset, a new dataset for machine learning techniques on MQTT},
  author={Vaccari, Ivan and Chiola, Giovanni and Aiello, Maurizio and Mongelli, Maurizio and Cambiaso, Enrico},
  journal={Sensors},
  volume={20},
  number={22},
  pages={6578},
  year={2020},
  publisher={MDPI}
}

@inproceedings{gil2016characterization,
  title={Characterization of encrypted and VPN traffic using time-related features},
  author={Gil, Gerard Drapper and Lashkari, Arash Habibi and Mamun, Mohammad and Ghorbani, Ali A},
  booktitle={Proceedings of the 2nd international conference on information systems security and privacy (ICISSP 2016)},
  pages={407--414},
  year={2016},
  organization={SciTePress Set{\'u}bal, Portugal}
}

@article{alzahrani2022artificial,
  title={Artificial intelligence algorithms for detecting and classifying MQTT protocol Internet of Things attacks},
  author={Alzahrani, Ali and Aldhyani, Theyazn HH},
  journal={Electronics},
  volume={11},
  number={22},
  pages={3837},
  year={2022},
  publisher={MDPI}
}

@Misc{ntop1,
author={ntop},
year={2023},
  title  = {How Flow-Based Traffic Classification Works},
  howpublished = {\url{https://www.ntop.org/how-flow-based-traffic-classification-works/}},
note= {Accessed: 23 November 2025} 
}

@Misc{zkDocProtocols,
author={Zeek},
year={2026},
  title  = {base/protocols/conn/main.zeek},
  howpublished = {\url{https://docs.zeek.org/en/master/scripts/base/protocols/conn/main.zeek.html}},
note= {Accessed: 04 June 2026} 
}

@Misc{zkDocBare,
author={Zeek},
year={2026},
  title  = {base/init-bare.zeek},
  howpublished = {\url{https://docs.zeek.org/en/lts/scripts/base/init-bare.zeek.html}},
note= {Accessed: 04 June 2026} 
}

@Misc{argusServerDebian,
author={debian Manpages},
year={2025},
  title  = {/ testing / argus-server / argus(8)},
  howpublished = {\url{https://manpages.debian.org/testing/argus-server/argus.8.en.html}},
note= {Accessed: 04 June 2026} 
}

@Misc{argusRscFile,
author={MANPAGES.ORG},
year={2025},
  title  = {argus resource file},
  howpublished = {\url{https://manpages.org/argusconf}},
note= {Accessed: 04 June 2026} 
}

@Misc{rashmi1,
author={Rashmi Bhardwaj},
year={2025},
  title  = {What is a 5-Tuple?},
  howpublished = {\url{https://ipwithease.com/what-is-a-5-tuple
}},
note= {Accessed: 23 November 2025} 
}

@Misc{argusweb,
author={argus},
year={2025},
  title  = {argus},
  howpublished = {\url{https://openargus.org/}},
note= {Accessed: 21 March 2025} 
}

@Misc{arborNetScout,
author={NETSCOUT},
year={2025},
  title  = {Arbor Sightline DDoS Attack Detection Solution
},
  howpublished = {\url{https://www.netscout.com/product/arbor-sightline}},
note= {Accessed: 21 November 2025} 
}

@Misc{ciscoStealth,
author={Cisco},
year={2025},
  title  = {Cisco Stealthwatch Enterprise Data Sheet
},
  howpublished = {\url{https://www.cisco.com/c/en_uk/products/collateral/security/stealthwatch/datasheet-c78-739398.pdf}},
note= {Accessed: 21 November 2025} 
}

@Misc{zeekorg,
author={Zeek},
year={2025},
  title  = {The Zeek Network Security Monitor},
  howpublished = {\url{https://zeek.org/}},
note= {Accessed: 21 March 2025} 
}

@inproceedings{poisson2024gothx,
  title={GothX: a generator of customizable, legitimate and malicious IoT network traffic},
  author={Poisson, Manuel and Carnier, Rodrigo and Fukuda, Kensuke},
  booktitle={Proceedings of the 17th Cyber Security Experimentation and Test Workshop},
  pages={65--73},
  year={2024}
}

@Misc{masteringMQTT,
author={EMQX},
year={2025},
  title  = {Mastering MQTT: The Ultimate Beginner's Guide for 2025},
  howpublished = {\url{https://www.emqx.com/en/blog/the-easiest-guide-to-getting-started-with-mqtt}},
note= {Accessed: 22 March 2025} 
}

@Misc{eqmx,
author={EMQX},
year={2025},
  title  = {Mastering MQTT-Your Ultimate Tutorial for MQTT
},
  howpublished = {\url{https://assets.emqx.com/resources/ebooks/Mastering\%20MQTT-Your\%20Ultimate\%20Tutorial\%20for\%20MQTT.pdf}},
note= {Accessed: 22 November 2025} 
}

@Misc{mqttattacktypes,
author={ Anthony Olazabal},
year={2024},
  title  = {Reinforcing Security of OT Systems in IIoT with MQTT and HiveMQ},
  howpublished = {\url{https://www.hivemq.com/blog/reinforcing-security-ot-systems-iiot-mqtt-hivemq/}},
note= {Accessed: 18 March 2025} 
}

@article{steinley2016variance,
  title={The variance of the adjusted Rand index.},
  author={Steinley, Douglas and Brusco, Michael J and Hubert, Lawrence},
  journal={Psychological methods},
  volume={21},
  number={2},
  pages={261},
  year={2016},
  publisher={American Psychological Association}
}

@article{garcia2026hybrid,
  title={Hybrid clustering-guided federated learning for robust intrusion detection in highly heterogeneous IoT environments},
  author={Garc{\'\i}a-S{\'a}ez, Luis Miguel and Ruiz-Villafranca, Sergio and Rold{\'a}n-G{\'o}mez, Jos{\'e} and Carrillo-Mond{\'e}jar, Javier and Mart{\'\i}nez, Jos{\'e} Luis},
  journal={Computer Networks},
  pages={112205},
  year={2026},
  publisher={Elsevier}
}

@article{rashed2021analysis,
  title={An analysis of Android malware classification services},
  author={Rashed, Mohammed and Suarez-Tangil, Guillermo},
  journal={Sensors},
  volume={21},
  number={16},
  pages={5671},
  year={2021},
  publisher={MDPI}
}

@inproceedings{diallo2021adaptive,
  title={Adaptive clustering-based malicious traffic classification at the network edge},
  author={Diallo, Alec F and Patras, Paul},
  booktitle={IEEE INFOCOM 2021-IEEE conference on computer communications},
  pages={1--10},
  year={2021},
  organization={IEEE}
}

@inproceedings{munz2007traffic,
  title={Traffic anomaly detection using k-means clustering},
  author={M{\"u}nz, Gerhard and Li, Sa and Carle, Georg},
  booktitle={Gi/itg workshop mmbnet},
  volume={7},
  number={9},
  pages={1159--35305},
  year={2007}
}

@article{baklashov2025statistical,
  title={Statistical and density-based clustering techniques in the context of anomaly detection in network systems: A comparative analysis},
  author={Baklashov, Aleksandr S and Kulyabov, Dmitry S},
  journal={Discrete and Continuous Models and Applied Computational Science},
  volume={33},
  number={1},
  pages={27--45},
  year={2025},
  publisher={Федеральное государственное автономное образовательное учреждение высшего~…}
}

@article{lo2025hierarchical,
  title={Hierarchical Clustering Approach With Distribution Similarity Strategy for Multi-Class Botnet Labeling in Real-World Traffic},
  author={Lo, Ta-Chun and Lin, Che-Hsien and Chang, Jyh-Biau and Shieh, Ce-Kuen},
  journal={IEEE Access},
  year={2025},
  publisher={IEEE}
}

@Misc{zeekflowmeter,
    author ={Github},
    year={2020},
  title  = {zkflowmtr},
  howpublished = {\url{https://github.com/zeek-flowmeter/zeek-flowmeter}},
note= {Accessed: 13 May 2025} 
}

@Misc{lyzerGithub,
      author ={Github},
    year={2024},
  title  = {ahlashkari / NTLFlowLyzer},
  howpublished = {\url{https://github.com/ahlashkari/NTLFlowLyzer}},
note= {Accessed: 04 June 2026} 
}

@Misc{mqttver5,
      author ={OASIS},
    year={2019},
  title  = {MQTT Version 5.0},
  howpublished = {\url{https://docs.oasis-open.org/mqtt/mqtt/v5.0/mqtt-v5.0.html}},
note= {Accessed: 20 March 2025} 
}

@Misc{amazQT,
      author ={Amazon},
    year={2025},
  title  = {What is MQTT?},
  howpublished = {\url{https://aws.amazon.com/what-is/mqtt/
}},
note= {Accessed: 22 November 2025} 
}

@article{husnain2022preventing,
  title={Preventing mqtt vulnerabilities using iot-enabled intrusion detection system},
  author={Husnain, Muhammad and Hayat, Khizar and Cambiaso, Enrico and Fayyaz, Ubaid U and Mongelli, Maurizio and Akram, Habiba and Ghazanfar Abbas, Syed and Shah, Ghalib A},
  journal={Sensors},
  volume={22},
  number={2},
  pages={567},
  year={2022},
  publisher={Multidisciplinary Digital Publishing Institute}
}

@inproceedings{hindy2020machine,
  title={Machine learning based IoT intrusion detection system: An MQTT case study (MQTT-IoT-IDS2020 dataset)},
  author={Hindy, Hanan and Bayne, Ethan and Bures, Miroslav and Atkinson, Robert and Tachtatzis, Christos and Bellekens, Xavier},
  booktitle={International networking conference},
  pages={73--84},
  year={2020},
  organization={Springer}
}

@article{vaccari2020slowite,
  title={SlowITe, a novel denial of service attack affecting MQTT},
  author={Vaccari, Ivan and Aiello, Maurizio and Cambiaso, Enrico},
  journal={Sensors},
  volume={20},
  number={10},
  pages={2932},
  year={2020},
  publisher={MDPI}
}

@inproceedings{ciklabakkal2019artemis,
  title={ARTEMIS: An intrusion detection system for MQTT attacks in Internet of Things},
  author={Ciklabakkal, Ege and Donmez, Ataberk and Erdemir, Mert and Suren, Emre and Yilmaz, Mert Kaan and Angin, Pelin},
  booktitle={2019 38th Symposium on Reliable Distributed Systems (SRDS)},
  pages={369--3692},
  year={2019},
  organization={IEEE}
}

@article{dadkhah2024ciciomt2024,
  title={Ciciomt2024: Attack vectors in healthcare devices-a multi-protocol dataset for assessing iomt device security},
  author={Dadkhah, Sajjad and Neto, E Carlos Pinto and Ferreira, Raphael and Molokwu, R Chukwuka and Sadeghi, Somayeh and Ghorbani, Ali},
 year={2024}
}

@Misc{adjustedRand,
  title  = {adjusted\_rand\_score},
        author ={Scikit-Learn},
    year={2025},
  howpublished = {\url{https://scikit-learn.org/stable/modules/generated/sklearn.metrics.adjusted_rand_score.html}},
note= {Accessed: 04 June 2026} 
}

@Misc{mqttHiveMQReport,
    author ={HiveMQ},
    year={2025},
  title  = {Expanding Industrial IoT in 2025
},
  howpublished = {\url{https://www.hivemq.com/resources/expanding-industrial-iot-in-2025-survey-report/}},
note= {Accessed: 22 February 2026} 
}

@Misc{mqttessentials,
    author ={HiveMQ},
    year={2025},
  title  = {MQTT Essentials},
  howpublished = {\url{https://www.hivemq.com/mqtt/}},
note= {Accessed: 27 July 2025} 
}

@misc{dachmanTCP2019,
  author       = {Dachman-Soled, D.},
  title        = {Attacks on TCP},
  year={2019},
  howpublished = {\url{https://user.eng.umd.edu/~danadach/Security_Fall_19/16_TCP_Attack.pdf}},
  note         = {Lecture notes for ENEE457/CMSC498E: Computer Systems Security, Fall 2019, University of Maryland. Accessed: 27 July 2025} ,
  year         = {2019}
}

@techreport{rfc7401,
  author = {R. Moskowitz, T. Heer, ...},
  title = {Host Identity Protocol Version 2 (HIPv2)},
  howpublished = {Internet Requests for Comments},
  type = {RFC},
  number = 7401,
  year = {2015},
  month = {04},
  issn = {2070-1721},
  publisher = {RFC Editor},
  institution = {RFC Editor},
  howpublished = {\url{https://datatracker.ietf.org/doc/html/rfc7401}}
}



\end{document}